\documentclass[manuscript]{aastex}
\shorttitle{Spitzer Mapping of PAH and H$_{2}$ features in Photodissociation Regions}
\shortauthors{Fleming et al.}
\usepackage{natbib}
\usepackage{graphicx}
\begin{document}
\title{{\bf $Spitzer$} Mapping of PAH and H$_{2}$ features in Photodissociation
  Regions}
\author{B. Fleming}
\affil{Department of Physics and Astronomy, Johns Hopkins University, Baltimore, MD 21218}
\email{flembri@pha.jhu.edu}
\author{K. France}
\affil{Center for Astrophysics and Space Astronomy, University of
  Colorado, Boulder, CO 80309}
\author{R. E. Lupu}
\affil{Department of Physics and Astronomy, University of Pennsylvania,
  Philadelphia, PA 19104}

\and 

\author{S. R. McCandliss} 
\affil{Department of Physics and Astronomy, Johns Hopkins University,
  Baltimore, MD 21218}

\begin{abstract}
The mid-infrared (MIR) spectra of dense photodissociation regions (PDRs) are typically dominated
by emission from polycyclic aromatic hydrocarbons (PAHs) and the
lowest pure rotational states of molecular hydrogen (H$_{2}$); two species which
are probes of the physical properties of gas and dust in
intense UV radiation fields. We utilize the high angular resolution of
the Infrared Spectrograph on the {\em Spitzer Space Telescope} to
construct spectral maps of the PAH and H$_{2}$ features for three of
the best studied PDRs in the galaxy, NGC 7023, NGC 2023 and IC
63. We present spatially resolved maps of the physical properties,
including the H$_{2}$ ortho-to-para ratio, temperature, and {\em
  G$_{o}$/n$_{H}$}. We also present evidence for PAH dehydrogenation,
which may support theories of H$_{2}$ formation on PAH surfaces, and a detection of preferential self-shielding
of ortho-H$_{2}$. All PDRs studied exhibit average
temperatures of $\sim$ 500~--~800~K, warm H$_{2}$ column densities of
$\sim$ 10$^{20}$ cm$^{-2}$, {\em G$_{o}$/n$_{H}$} $\sim$ 0.1~--~0.8,
and ortho-to-para ratios of $\sim$ 1.8. We find that while the average of each of these
properties is consistent with previous single value measurements of these PDRs, when available, the
addition of spatial resolution yields a diversity of values with gas temperatures as high as
1500 K, column densities spanning $\sim$ 2 orders of magnitude, and
extreme ortho-to-para ratios ({\em R$_{OP}$}) of {\em R$_{OP}$}
  $<$ 1 and {\em R$_{OP}$} $>$ 3.
\end{abstract}


\section{Introduction}

Photodissociation regions (PDRs) around hot stars are the boundaries between the
ionized HII region and the surrounding molecular clouds. PDRs consist of layers of atomic and molecular emission with a
chemical profile dictated by the ratio {\em G$_{o}$/n$_{H}$}, where
{\em G$_{o}$} is the strength of the local UV radiation field
(912~--~2000~\AA ) in units of the Habing flux (1.6 $\times$ 10$^{-3}$ erg
cm$^{-2}$ s$^{-1}$, \citet{HAB68}) and {\em n$_{H}$} is the local
hydrogen density \citep{HTI97}. The typical mid-infrared (MIR; 5-35 $\mu$m) spectrum of a PDR is dominated by the stretching
and bending modes of polycyclic aromatic hydrocarbons (PAHs) excited
by the absorption of individual UV photons and by the
lowest pure
rotational lines of molecular hydrogen \citep{LAP84}. H$_{2}$ is formed
on the surface of dust grains and is destroyed by
photodissociation within PDRs. Its formation rate has been found to be
related to PAH abundance \citep{HAB04}, suggesting that there may be
an observational relationship
between these two species. Both PAHs and H$_{2}$ have
been shown to be valuable probes of the physical conditions within
PDRs \citep{BAT94,DAB96,HTI97,BRN09}. 

PAH emission in the MIR is dominated by the five broad emission features centered at 6.2
$\mu$m, 7.7 $\mu$m, 8.6 $\mu$m, 11.3 $\mu$m and 12.7 $\mu$m. These,
along with features at 3.3 $\mu$m and 17 $\mu$m, are also referred to
as the Unidentified Infrared (UIR) features and the Aromatic Infrared
Bands (AIBs). Studies have suggested that each of the major PAH
bands is actually a blend of emission from PAHs of various sizes,
surface structures, and charges, therefore the relative shapes,
centroids and flux ratios with respect to the other bands
vary depending on the physical properties of the observed PAHs \citep{BAT94,PEET02,VanDieden04}. Laboratory measurements have shown that the
features attributed to carbon-hydrogen (C-H) bond bending, namely the 3.3
$\mu$m, 11.3 $\mu$m and 12.7$\mu$m bands, are dominant for a
population of neutral PAHs, however the carbon-carbon (C-C) stretching modes, from
6 - 9 $\mu$m, are the dominant emission from PAH cations (PAH$^{+}$)
\citep{ALA99,HUDGE99}. In recent years the ratio of C-C emission to
C-H emission, most notably the 6.2 $\mu$m or 7.7 $\mu$m to 11.3 $\mu$m
bands, has become a method of estimating the charge state of a PAH
cloud \citep{JOB96,GAL08}. A smaller number of studies have shown that the ratio of the 12.7 $\mu$m emission, which
arises from duo and trio C-H out-of-plane bending modes, to the 11.3
$\mu$m singlet C-H out-of-plane bending mode is a probe of the PAH surface structure
\citep{DuleyWilliams81,Hony01,TIE08}. For smaller PAHs (with fewer
than 50 carbon atoms), {\em I$_{12.7}$/I$_{11.3}$} may be an indicator of
dehydrogenation, or the breakdown of duo and trio C-H bonds into solo
bonds.  

\begin{figure}
\plotone{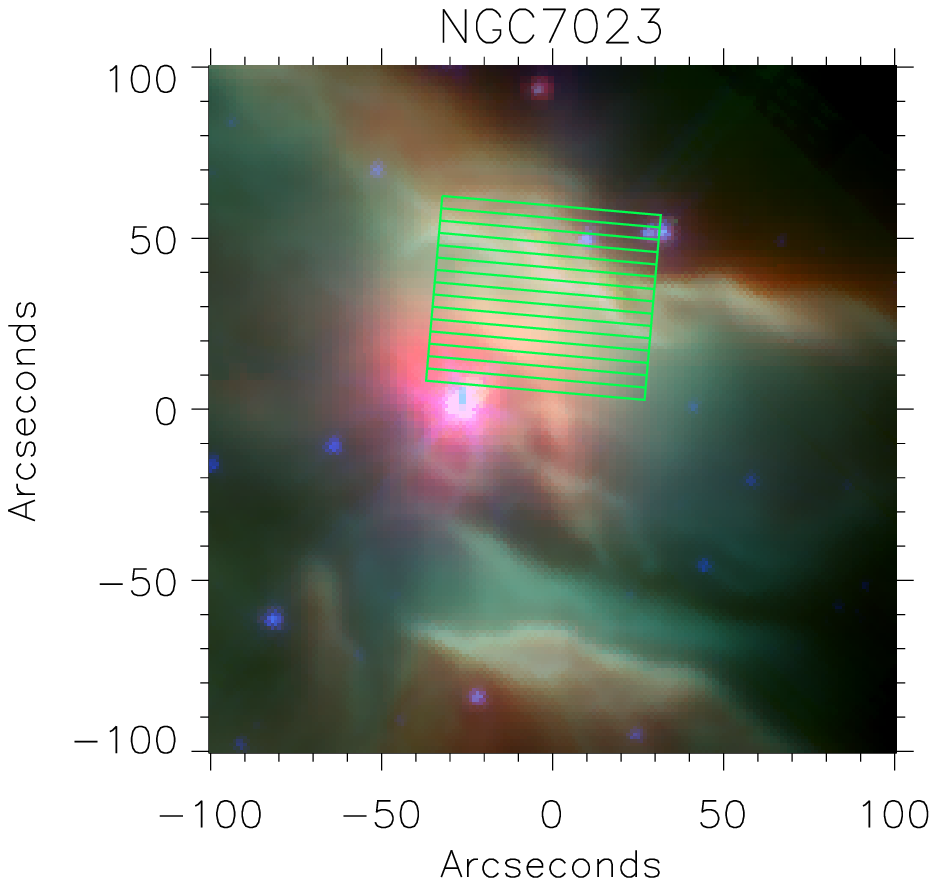}
\caption{IRAC 3.6$\mu$m, 8$\mu$m, and MIPS 24$\mu$m image of NGC7023 with IRS slit
  overlays. Overlapping slits are merged. North is oriented up in this image. \label{fig-7023}}
\end{figure}

\begin{figure}
\plotone{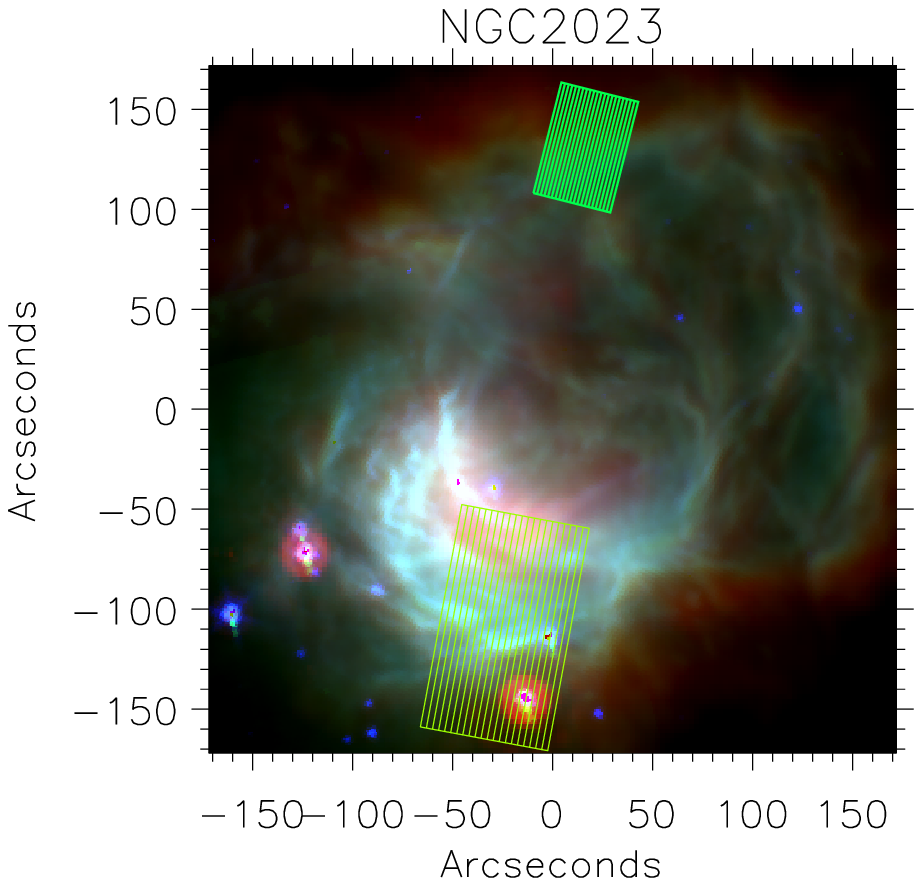}
\caption{IRAC 3.6$\mu$m, 8$\mu$m, and MIPS 24$\mu$m image of NGC2023 with IRS slit
  overlays. Overlapping slits are
  merged. North is oriented up in this image. \label{fig-2023}}
\end{figure}

\begin{figure}
\plotone{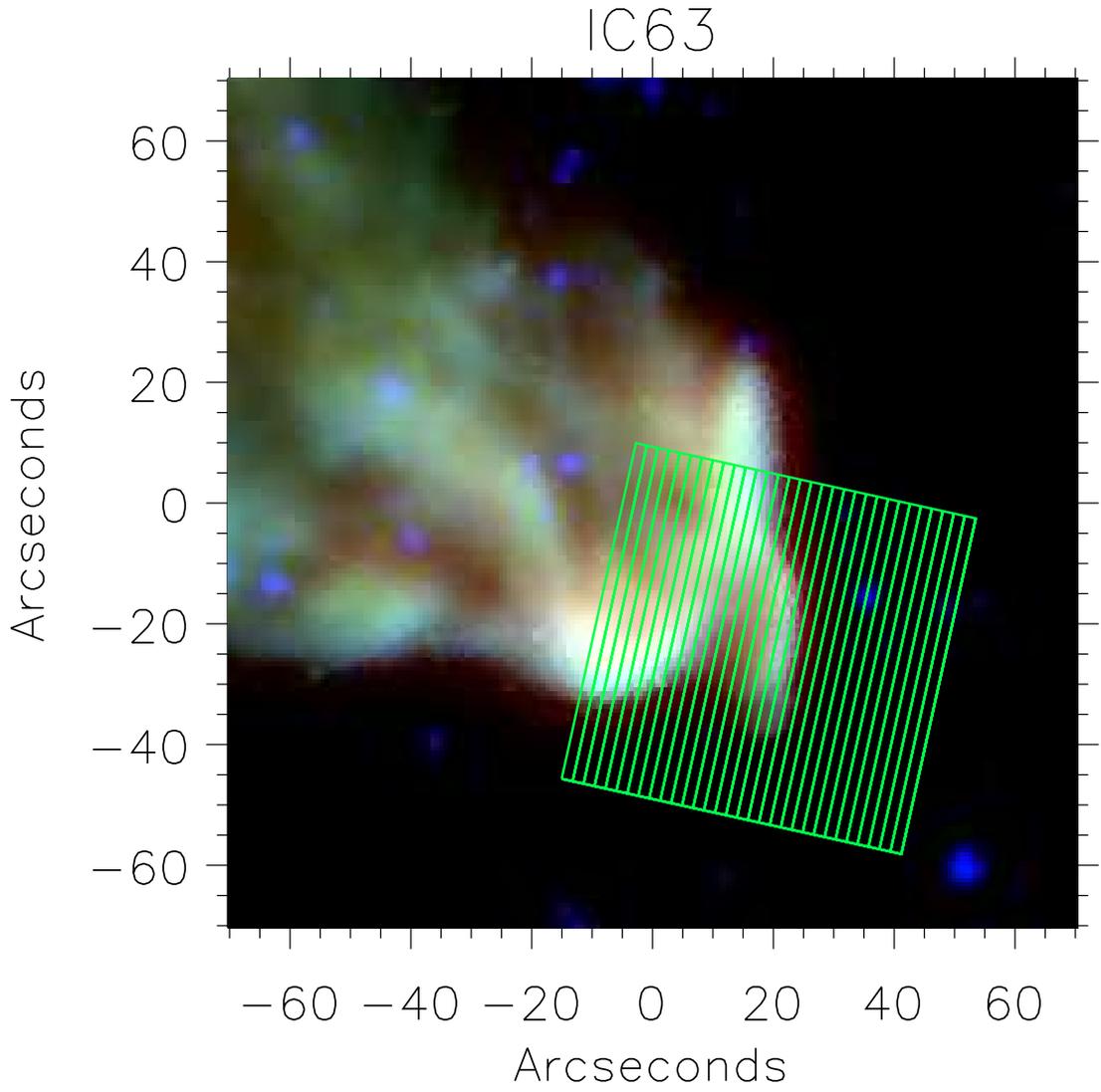}
\caption{IRAC 3.6$\mu$m, 8$\mu$m, and MIPS 24$\mu$m image of IC63 with IRS slit
  overlays. The star $\gamma$ Cas is approximately 1.3 pc SW of this
  emission nebula and not in this image. North is oriented up. \label{fig-63}}
\end{figure}

In this paper, we present spectral line maps of the H$_{2}$ and PAH
emission measured by the {\em Spitzer Space Telescope} from four of the best studied PDRs; NGC 7023, IC 63 and two
separate regions in NGC 2023 (Figures \ref{fig-7023}, \ref{fig-2023},
\ref{fig-63}). All three PDRs are known to exhibit UV
fluorescence of H$_{2}$ (\citet{FRA09} and references therein). We utilize the high angular resolution
of the Infrared
Spectrograph (IRS; \citet{HOU04}) onboard {\em Spitzer} to investigate the spatial variation of
the temperature, column density, and ortho-to-para ratio of the lowest pure rotational
states of H$_{2}$, the ratio {\em G$_{o}$/n$_{H}$} determined from the
ionization state of the PAHs, and we address the possibility of dehydrogenation
of PAHs across the observed regions. We present
these maps and the methods used to generate them in $\S$3. In $\S$4 we
present a summary of the new information revealed by the mapping
technique for each PDR, and address the agreement between our
measurements and previous observations. In $\S$5 we address and similarities or
differences between the PDRs and summarize our findings. A summary of the
  basic properties of the regions is presented
  in Table \ref{tbl-prop}.

\section{Observations and Data Reduction}
\subsection{Data Cube Assembly}
{\it Spitzer} data products were downloaded from the
{\it Spitzer} Science Center (SSC) archive for NGC 7023, IC 63, and two
locations in NGC 2023: NGC 2023 North (NGC 2023N), located $\sim$
160\arcsec\ north of the central star HD 37903, and NGC 2023 South (NGC
2023S), located $\sim$ 78\arcsec\ south of HD 37903. {\em IRS} short wavelength low resolution (R $\sim$ 60 - 127,
SL) 5 - 14 $\mu$m
Basic Calibrated Data (BCD) spectra were obtained for each
nebula. For a portion of NGC 7023 and for all of NGC 2023N we also obtained long wavelength low resolution (R $\sim$ 57 -
126, LL) 14 - 38 $\mu$m spectra. These spectra were assembled into data
cubes using the CUBISM software \citep{JDS07}

{\it Spitzer} BCD products require background and bad pixel
subtraction. In IC 63, the outrigger exposures (the exposures of
channel not centered on the target) were clear of nebulosity
and therefore supplied a suitable background exposure. This
was not the case for the extended PDRs NGC 7023 and NGC 2023. For NGC
2023S a dedicated sky exposure was taken and included in the BCD data
package. To account for background emission in NGC 7023 and NGC 2023N, we searched the
SSC archive for point source spectra of objects within 10$\degr$ of
declination of each of our targets taken within $\pm$ 3 days of our
data. If the outrigger orders of these exposures were free of stars
or nebulosity, then it was considered suitable for background subtraction. 

Bad pixels were removed using the CUBISM
automated bad pixel routine as well as by interactive selection. Mapping
observations with overlapping slits along the slit axis required
editing of the default 'WAVSAMP' to eliminate contamination caused by
the higher noise at the slit endpoints. The resulting 3-D data cubes
(two spatial dimensions and one spectral, N $\times$ M $\times$ $\lambda$) from each of the two IRS channels were merged into one by averaging
the regions of overlap between each of the spectral orders (the bonus orders were
ignored). The spatial dimensions of the data cubes are $\sim$ 1.8
$\times$ 1.8 $\arcsec$/pixel. A
summary of the {\it Spitzer} AORs, program ID's and background program
IDs for each of our datasets is listed in Table \ref{tbl-obs}. 

\subsection{Spectral Feature Fitting}
Each 1-D spectrum in the data cube was fit
using the PAHFIT IDL routine \citep{JDS07}. PAHFIT uses Drude
profiles to fit the broad PAH features and Gaussians to fit the atomic and
molecular lines, including the H$_{2}$ 0-0 S(0) -- S(7) rotational lines. The use of Drude profiles for the aromatic bands has
been shown to recover $\sim$2-6 times more flux in the broad wings of
the 7.7 $\mu$m and 8.6 $\mu$m features than
spline fitting or Gaussian profiles \citep{JDS06}. The resulting
integrated fluxes for each feature were then reassembled into maps. 

PAHFIT has a built-in routine to correct for extinction, including the
silicate absorption features at 10 and 18 $\mu$m.  To improve the spatial correlation of the extinction
fitting, we re-binned our spectral maps into larger, 18 $\times$
18$\arcsec$ pixels ($\sim$ 5 $\times$ 5 IRS pixels). The resulting extinction values were
extrapolated, smoothed and then applied to the original maps using the
PAHFIT extinction curve. For the majority of our sample the MIR extinction is nearly negligible
(the optical depth at 9.7 $\mu$m, $\tau_{9.7} <$ 0.2), which is
consistent with the {\em ISO} derived global value inferred from observation
of HD 200775 for NGC 7023 \citep{FUE00}, and also with extrapolations
of optical and near-IR measured extinctions into the MIR \citep{DAB00}. 

The data for NGC 2023N has an anomalous
absorption feature between 9.3 and 10.7 $\mu$m with a position dependent
centroid. The average full width at half maximum of the feature is $\sim$
0.32 $\mu$m and the average integrated flux is $\sim$ 27.25 MJy
$\mu$m sr$^{-1}$.  The centroid varies between $\sim$ 9.25 $\mu$m and
10.65 $\mu$m.  We attempt to correct for by first determining the
centroid and then scaling the original spectra
over that region using a multi-gaussian fit and the average FWHM and
integrated flux listed above.  This correction introduces a systematic uncertainty to the H$_{2}$ S(3) rotational
line. Due to the importance of S(3) as the primary SL band
ortho-rotational line detected in NGC 2023N, we utilize S(3) in our
analysis despite the uncertainty. We estimate the error in the S(3)
flux to vary from $\sim$ 0\% near the edges of the observed region to
up to 40\% in the middle, however we see no position dependant
artifacts in our temperature, column density, or H$_{2}$ ortho-to-para
ratio maps ($\S$3).  Nevertheless, this ad hoc correction should be taken into account
when considering the validity of our results for that PDR.

\section{Spectral Maps}
\begin{figure}
\epsscale{0.79}
\plotone{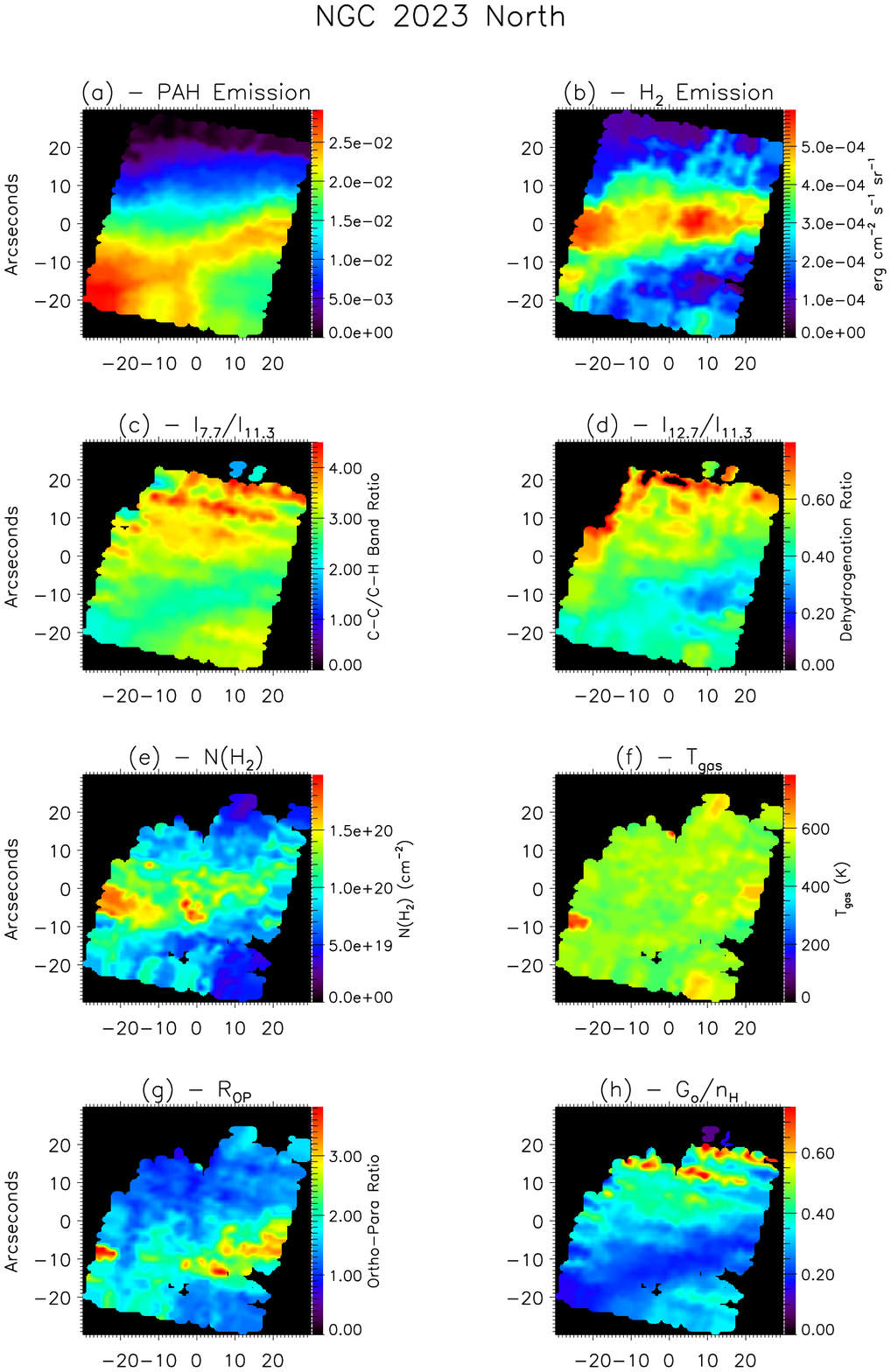}
\caption{PAH and H$_{2}$ emission and diagnostics for NGC 2023 North. (a)
  Total detected PAH emission. (b) Total detected H$_{2}$
  emission. (c) {\em I$_{7.7}$/I$_{11.3}$}. (d) {\em
    I$_{12.7}$/I$_{11.3}$}. (e) H$_{2}$ column denisty. (f) H$_{2}$
  derived gas temperature. (g) H$_{2}$ Ortho-to-para ratio. (h) {\em G$_{o}$/n$_{H}$}.
\label{fig-2023Na}}
\end{figure}

\begin{figure}
\epsscale{0.79}
\plotone{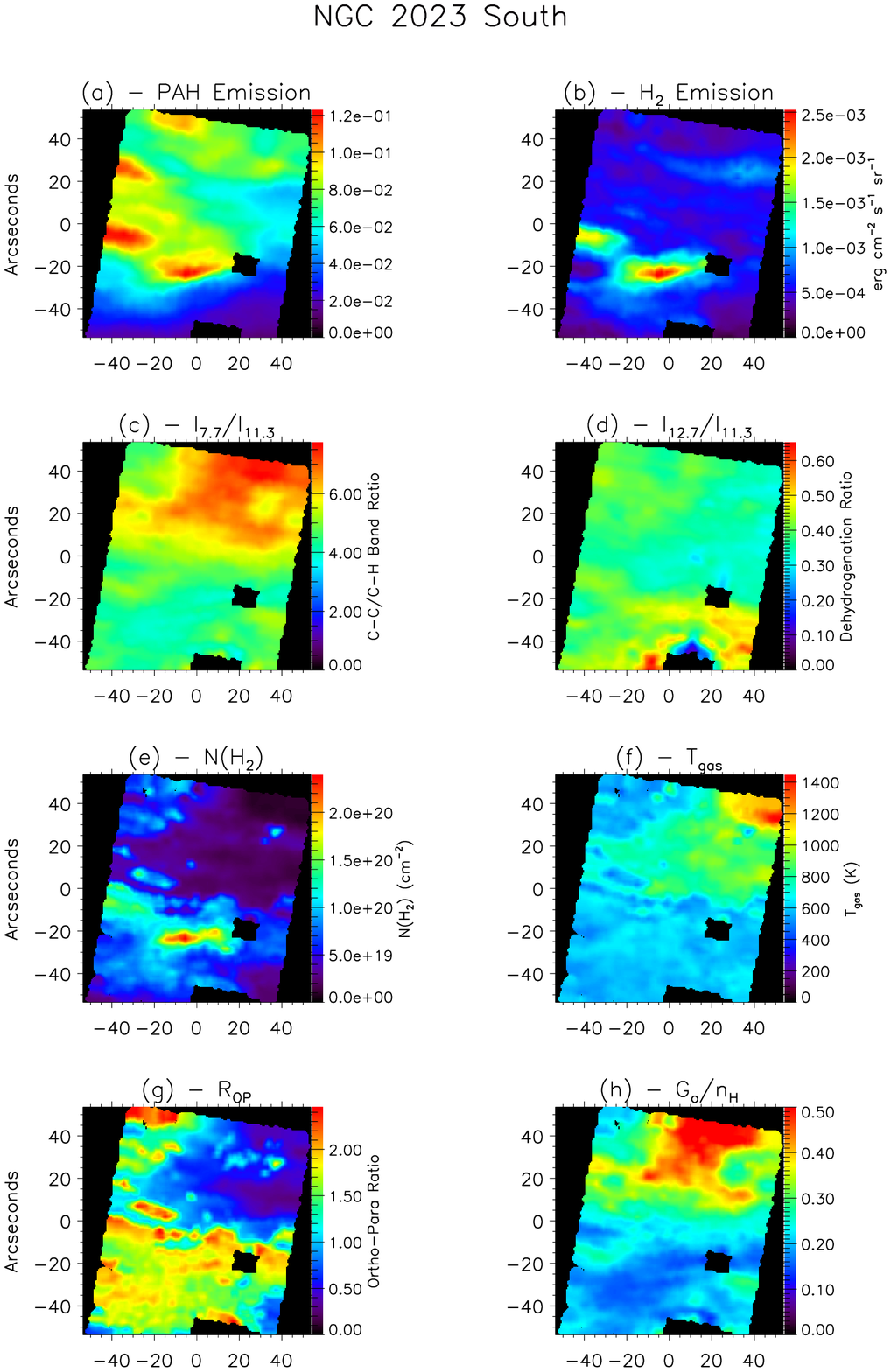}
\caption{PAH and H$_{2}$ emission and diagnostics for NGC 2023 South. (a)
  Total detected PAH emission. (b) Total detected H$_{2}$
  emission. (c) {\em I$_{7.7}$/I$_{11.3}$}. (d) {\em
    I$_{12.7}$/I$_{11.3}$}. (e) H$_{2}$ column denisty. (f) H$_{2}$
  derived gas temperature. (g) H$_{2}$ Ortho-to-para ratio. (h) {\em G$_{o}$/n$_{H}$}.
\label{fig-2023Sa}}
\end{figure}

Spectral maps of the total H$_{2}$ and PAH emission and the resulting
line ratios and diagnostics presented in this section are included in Figures \ref{fig-7023a}
-- \ref{fig-63a}.  Figures \ref{fig-7023a} -- \ref{fig-63a} (a) and (b)
represent the total detected SL band PAH and H$_{2}$ emission
respectively. Figures \ref{fig-7023a} -- \ref{fig-63a} (c) and (d)
represent the ratio of the 7.7 $\mu$m to 11.3 $\mu$m and 12.7 $\mu$m
to 11.3 $\mu$m PAH bands respectively. Figures \ref{fig-7023a} --
\ref{fig-63a} (e), (f), and (g) represent the calculated H$_{2}$
column densities, temperatures, and ortho-to-para ratios. Figures \ref{fig-7023a} -- \ref{fig-63a} (h) show the
calculated values of {\em G$_{o}$/n$_{H}$}. In the following section
we outline how each of these maps is constructed and discuss their
physical interpretation.\footnote{Maps of each individual H$_{2}$ rotation line, the major PAH features,
and properties calculated in this paper are available online at
{\tt www.pha.jhu.edu/$\sim$flembri/PDRs}.}

For consistency, spectral analysis is only performed on data from the IRS SL channel, except where noted. Therefore when we refer to PAH
emission, we are neglecting the PAH features around 17 $\mu$m, and when
we refer to H$_{2}$ emission, we are neglecting S(0) and S(1). All
maps have been oriented so that north is up. A summary of the average
values for each of the properties calculated is presented in Table
\ref{tbl-avgs}. These average values are not the same as the values
obtained if the entire observed region was rebinned into a single
pixel, however, as equal weight is given to each pixel regardless
of the flux of the lines used to calculate that property. Therefore, the
temperature of a pixel with an order of magnitude lower H$_{2}$ flux
than another is counted the same when computing the
average, even though the pixel with the higher flux would dominate in
a larger aperture. Single value measurements are already available for these
PDRs and listed in Table \ref{tbl-prop}. 

\subsection{PAH and H$_{2}$ Emission}
Individual spectral maps were constructed for the major {\em Spitzer}
SL band PAH features (6.2, 7.7, 8.6, 11.3, and 12.7 $\mu$m) and H$_{2}$ 0--0 S(J = 2 -- 7)
rotational emission lines using the extinction corrected line fluxes output by the PAHFIT
routine. The signal-to-noise of the integrated line flux extracted from each 1-D spectra was required to be $\geq$ 3 for the line to be included in its spectral map. Figures \ref{fig-7023a}
-- \ref{fig-63a} (a) and (b) represent the sum of the individual PAH
and H$_{2}$ maps respectively and therefore are maps of the total {\em detected}
PAH and H$_{2}$ flux in the 5 -- 14 $\mu$m bandpass.

The individual PAH features are strong,
especially in NGC 7023 and NGC 2023, and Figures \ref{fig-7023a}
-- \ref{fig-63a} (a) represent the sum of the individual spectral maps
with nearly complete coverage of the observed PDR in all
bands. Figures \ref{fig-7023a} -- \ref{fig-63a} (b), however, only represent a lower limit on the total SL
band H$_{2}$ emission as the signal-to-noise of the higher {\em J} lines
is often below our S/N $\geq$ 3 threshold. Throughout this paper we neglect the H$_{2}$ S(6) (6.11 $\mu$m) line as it is heavily blended with the
stronger 6.2 $\mu$m PAH feature and therefore cannot be reliably
extracted. 

\subsection{PAH Band Ratios}
In recent years measurements of the PAH ionization state have
become important diagnostic tools for PDRs. \citet{BRN07} successfully
isolated PAH$^{0}$
and PAH$^{+}$ emission using blind signal separation techniques,
showing that PAH$^{+}$ emission is most prevalent closer to the
HII region where the radiation field is stronger. We investigate the degree of PAH
ionization using the ratio of the 7.7 $\mu$m C-C
stretching modes to the 11.3 $\mu$m C-H out-of-plane bending
mode. Laboratory studies have shown that ionized PAHs emit more in the
C-C modes relative to the C-H bending modes than neutral PAHs,
therefore Figures \ref{fig-7023a} -- \ref{fig-63a} (c)
are representations of the ionization state of the PAH molecules in
the PDRs
\citep{JOB96, ALA99}.  The ratio {\em I$_{7.7}$/I$_{11.3}$} is
expected to be $\sim$ 1.3 for a fully neutral column of PAHs and
  $\sim$ 12 for a completely ionized column \citep{LID01}. 

 The maps presented in Figures
\ref{fig-7023a} -- \ref{fig-63a} (d) represent the surface structure of the PAHs given by the ratio {\em I$_{12.7}$/I$_{11.3}$}. PAHs can be stripped down to their carbon cores, or
even dissociated via carbon fragment loss, by absorbing UV photons
with energies as low as 5 -- 8 eV \citep{LePage03}. The
dehydrogenation of a PAH decreases emission in the C-H singlet bands
(centered at 11.3 $\mu$m) relative to the duo and triplet bands
(centered at 12.7 $\mu$m), however this relation is not fully understood, therefore we will focus only on what the trends in {\em I$_{12.7}$/I$_{11.3}$} reveal rather than the absolute values of
the ratios. 

\subsection{H$_{2}$ Temperature, Column Density, and Ortho-to-Para Ratio}
We can determine a gas temperature, {\em T$_{gas}$}, using the pure rotational states of
H$_{2}$. These transitions have small critical densities and therefore can be
assumed to be dominated by collisional de-excitation (see chap. 9.7.3 of
Tielens (2005) for a review). Assuming thermodynamic equilibrium and a
Maxwell-Boltzmann distribution,

\begin{equation}
\ln \bigg(\frac{N_{J+2}}{g_{J+2}}\bigg)=\ln{(N_{H_{2}})}-\frac{E_{J+2}}{kT_{gas}} ,
\end{equation}
where E$_{J+2}$ is the energy and {\em g$_{J+2}$} is the statistical weight of the
{\em J+2} state. H$_{2}$ is a homonuclear molecule and hence has no
dipole allowed pure rotational transitions, therefore we only detect quadrupole transitions
with {\em $\Delta$J} = 2. The statistical weight of the {\em J}
state is,

\begin{equation}
\begin{array}{l}
\displaystyle g_{J=0,2,4,6...} = 2J+1 \\
\displaystyle g_{J=1,3,5,7...} = (2J+1)R_{OP} \\
\end{array} ,
\label{eq:gs}
\end{equation}
where {\em R$_{OP}$} is the ortho-to-para ratio of the H$_{2}$, which
is 3 in a gas at temperature greater than $\sim$ 250 K in local
thermal equilibrium (LTE) \citep{SNU99}.  For H$_{2}$ out of LTE, the
ortho-rotational lines (odd {\em J} states corresponding to parallel
nuclear spins) will be offset from the para-rotational
lines (even {\em J} states with anti-parallel spins), resulting in a ``zig-zag''
distribution.  The column density, N$_{J+2}$, of the {\em J}+2 state is
determined by the Einstein {\em A$_{S(J)}$} coefficient,

\begin{equation}
N_{J+2} = \frac{4\pi f_{S(J)}}{A_{S(J)}  hc /\lambda_{S(J)}},
\end{equation}
where S(J) denotes the H$_{2}$ 0-0 ({\em J} + 2) $\rightarrow$ {\em J}
rotational transition and {\em f$_{S(J)}$} is the measured flux density of the rotational emission
line. This equation is valid only because the H$_{2}$ lines
are dipole-forbidden and therefore optically thin. 

In a uniform, single temperature gas with {\em R$_{OP}$} = 3, the curve
established by Equation [1] will be a straight line on a
log-linear plot. The slope of this line is defined by the H$_{2}$
rotational temperature. In a single temperature gas, the temperature, column density, and
{\em R$_{OP}$} can be determined by iterating {\em R$_{OP}$} until the
data can be fit by a single {\em N$_{H_{2}}$} and
{\em T$_{gas}$}.    

\begin{figure}
\plotone{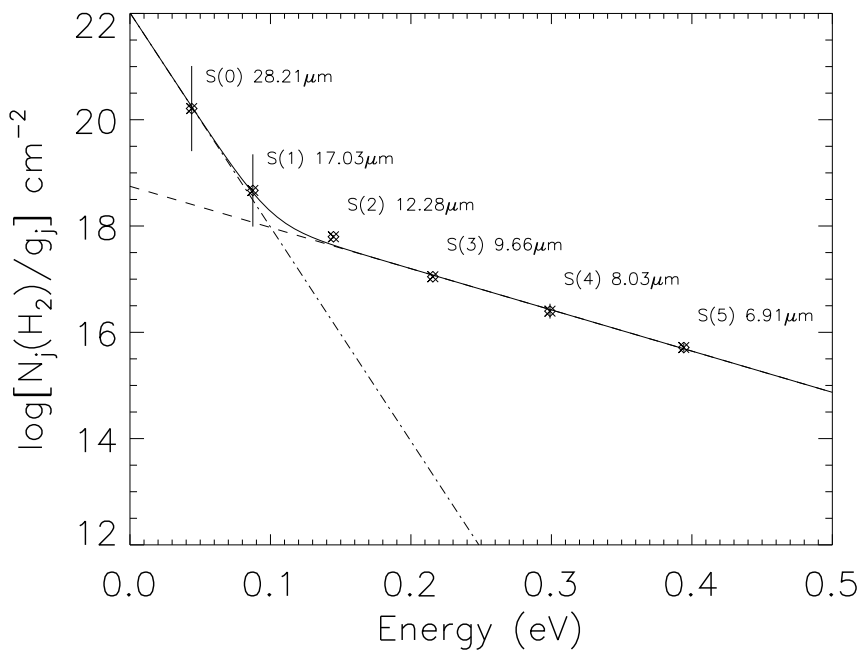}
\caption{Plot of the H$_{2}$ S(0) - S(5) rotational lines as a
  function of transition energy for a single data cube element near the bright ribbon of
  NGC 7023. Overplotted are two temperature fits, one for a higher
  column density gas (3.5 $\times$ 10$^{22}$ cm$^{-2}$) at 125K (dot-dash)
  and another for a lower column density gas (9 $\times$ 10$^{19}$ cm$^{-2}$) at 650K (dash). 
\label{fig-tempcurv}}
\end{figure}

PDRs consist of a multiphase gas that can be approximated by a low
density, diffuse medium interspersed with high density clumps or
filaments \citep{CKS88,Sellgren92}. The majority of the
H$_{2}$ is in the cool, low density phase, however the emission
spectrum in the SL band is dominated by the higher density warm
component, often making up $<$ 1\% of the total H$_{2}$ column (Figure
\ref{fig-tempcurv}).  For the purposes of this study, we assume a
single temperature dominates the H$_{2}$ S(2) -- S(5) rotational
emission lines and fit a temperature, {\em N$_{H_{2}}$} and {\em
  R$_{OP}$} accordingly. To construct maps of these properties (Figures \ref{fig-7023a} 
-- \ref{fig-63a} (e), (f) and (g) respectively), we require that at least three
of the four H$_{2}$ S(2) -- S(5) rotational emission lines have a
signal-to-noise ratio greater than 3 and that the line established by
the fit falls within the 1$\sigma$ error bars of all three of the measurements. Due to the lack of LL coverage in our sample
and the low statistical weight of the S(0) line,
establishing the temperature, {\em N$_{H_{2}}$}, and {\em R$_{OP}$} of the cool H$_{2}$ is not
possible except in NGC 2023N and a limited portion of NGC 7023.  

\subsubsection{H$_{2}$ S(4) Overestimation}
In NGC 7023 and NGC 2023, H$_{2}$ S(4)
consistently falls above the fit defined by S(2), S(3) and S(5). We determined that this is due to the S(4) line, at
8.06 $\mu$m, overlapping the broad 7.7 $\mu$m PAH feature. The H$_{2}$
features in both of these PDRs are weak compared to the PAH
emission (see Figure \ref{fig-7023a} -- \ref{fig-2023Sa} (a) and (b))
and the PAHFIT routine overestimates the S(4) flux in these two PDRs
by $\sim$ 10-50$\%$. This error causes an overestimation of the
H$_{2}$ temperature of $\sim$ 30 K and N(H$_{2}$) of $\sim$ 10$\%$.  For
each PDR, we used our temperature fit to the S(2), S(3) and S(5) lines to estimate
the actual S(4) flux and then applied the average overestimation,
relative to the PAH 7.7 $\mu$m flux, as a correction to the S(4) flux
for the entire PDR. This correction was necessary as the S(5) line
is weaker than the S(4) line in these PDRs and not always detected. 

The ortho-to-para ratio is much more sensitive to the S(4)
overestimation than the temperature or column density. While our
correction does improve the quality of the {\em R$_{OP}$} fit, it also introduces an additional degree of
uncertainty, therefore we do not use
H$_{2}$ S(4) to estimate ortho-to-para ratios. We utilize all S(2) -
S(5) rotational lines to fit a temperature (which simultaneously fits
an ortho-to-para ratio) and then we remove the S(4) line, set the
temperature and column density, and determine the best fit
ortho-to-para ratio to the remaining lines. This significantly reduced the scatter of
the ortho-to-para maps (Figures \ref{fig-7023a} --
\ref{fig-2023Sa} (g)). IC 63, which has lower total PAH flux
relative to H$_{2}$ emission (Figures \ref{fig-63a} (a) and (b)), does
not show evidence of H$_{2}$ S(4) overestimation.   

\subsection{G$_{o}$/n$_{H}$}
The gas temperature and the ionization state of PAHs can be used as a probe of the local density and UV
field strength, {\em G$_{o}$/n$_{H}$}. Using the parameter
fit for the 11.3 $\mu$m flux ratio of
ionized to neutral PAHs ({\em I$^{+}_{11.3}$/I$^{0}_{11.3}$} $\sim$ 0.6) and Equation [2] from \citet{JOB96},

\begin{equation}
\frac{I_{7.7\mu m}}{I_{11.3\mu m}} = 1.3 \bigg(\frac{1+5.54 \frac{PAH^{+}}{PAH^{0}}}{1+0.6
  \frac{PAH^{+}}{PAH^{0}}} \bigg) ,
\end{equation}
where {\em PAH$^{+}$} and {\em PAH$^{0}$} represent the total number of ionized
and neutral PAHs respectively. This equation can be related to {\em $\sqrt{T_{gas}}$G$_{o}$/n$_{e}$} by solving for the neutral fraction,
{\em f$_{n}$} = {\em PAH$^{0}$/(PAH$^{+}$+PAH$^{0}$)}, and setting it equal to the ionization
parameter relation, {\em f$_{n}$} = {\em (1+$\gamma_{o}$)$^{-1}$} \citep{BAT94,TIE05}. 

\begin{equation}
\gamma_{o}=3.5 \times 10^{-6} N_{C}^\frac{1}{2}\frac{G_{o}T^{1/2}}{n_{e}},
\end{equation}

\begin{equation}
\bigg(\frac{I_{7.7}/I_{11.3}-1.3}{7.2-0.6(I_{7.7}/I_{11.3})}\bigg) \approx
3.5x10^{-6}N_{C}^{\frac{1}{2}}\bigg(\frac{G_{o} \sqrt{T_{gas}}}{n_{e}}\bigg).
\end{equation}
We adopt an average number of carbons atoms per PAH of
{\em N$_{C}$} = 50. This approximation breaks down near the total ionization
limit of {\em I$_{7.7}$/I$_{11.3}$} $\rightarrow$ 12, however we do not encounter
{\em I$_{7.7}$/I$_{11.3}$} $>$ 11
except in NGC 7023 in the inner nebular regions near HD 200775. We
do not include data from that region in the following
analysis. We estimate the electron density, {\em n$_{e}$} by assuming that all
free electrons come from singly ionized gaseous carbon, and that all atomic carbon in the
PAH emission region is ionized. Using the C/H ratio from \citet{SOF04}, the electron density is
estimated to be {\em n$_{e}$} $\approx$ 1.6 $\times$ 10$^{-4}$ n$_{H}$. Maps of {\em G$_{o}$/n$_{H}$} are presented in Figures
\ref{fig-7023a} (h) -- \ref{fig-63a} (h). 

\section{Results}
\subsection{NGC 7023}
\begin{figure}
\plotone{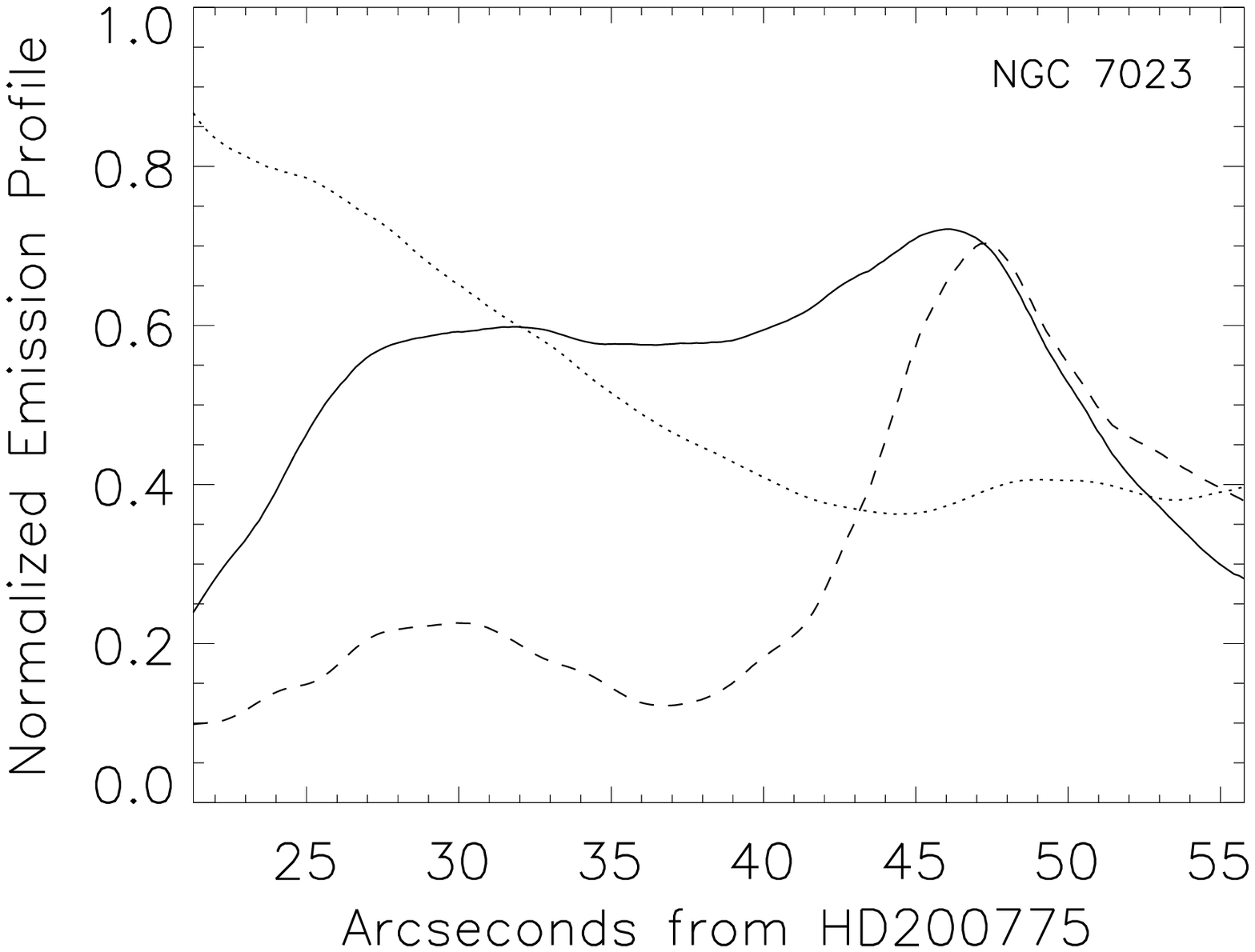}
\caption{Average normalized emission profile for PAHs (solid), H$_{2}$
  (dash) and (I$_{7.7}$/I$_{11.3}$)/12 (dot) for NGC 7023. Multiple
  cross-sections through the PDR originating from the direction of HD 200775 are averaged to produce this profile.  
\label{fig-7023cut}}
\end{figure}

In Figures \ref{fig-7023a} (a) and (b) it is clear that the PAHs and
H$_{2}$ do not share the same morphology. The PAHs extend $\sim$
20$\arcsec$ closer to HD 200775 than the H$_{2}$, which rapidly falls
from its peak at the starwards edge of the ribbon feature. Figure
\ref{fig-7023cut} shows that the PAH emission peaks on average $\sim$ 1.5$\arcsec$ starwards of the
H$_{2}$. The separation of H$_{2}$ rotational
emission and PAHs has been shown before, most notably across the
Orion Bar \citep{TIE93,Kassis06}, however in PDRs with
weaker UV fields this has not been observed \citep{COMP07}.  

\subsubsection{PAH Band Ratios}
Figure \ref{fig-7023a} (c) shows that the PAH band ratio {\em I$_{7.7}$/I$_{11.3}$} is at a minimum near the
bright ribbon feature and increases with decreasing projected distance to HD
200775. This is consistent with the findings of \citet{RAPA05}, who
used {\em ISO} spectra to show that the relative contribution to the
PAH spectrum of ionized PAHs increases closer to the central
star. The geometry of the band ratio map does not match any of the
other MIR observables and we note no apparent correlation between
total PAH
emission and ionization until {\em I$_{7.7}$/I$_{11.3}$} $\sim$ 8. At
this ratio, which from Equation [4] indicates an ionization fraction
of {\em f$_{+} \sim$} 0.74, the total PAH flux begins to rapidly
decline, potentially indicating destruction, falling by
more than half over $\sim$ 5$\arcsec$
(Figure \ref{fig-7023cut}). 

The dehydrogenation ratio {\em
  I$_{12.7}$/I$_{11.3}$} is highest both in the bright ribbon feature
and close to HD 200775 (Figure \ref{fig-7023a} (d)). Beginning at
the starwards edge of the dissociation front and extending $\sim$
10$\arcsec$ towards the central star there is a trough where the ratio
reaches a minimum. As with the ionization ratio, the geometry
does not follow that of any other observable.   

\subsubsection{H$_{2}$ Features}
The increase in H$_{2}$ column density (Figure \ref{fig-7023a} (e)) and flux at the bright
ribbon feature is likely due to the onset of H$_{2}$ self-shielding,
which is a rapid process. In the ribbon feature the {\em R$_{OP}$} of
the H$_{2}$ is a nonequilibrium value of $\sim$ 1.83 $\pm$ 0.51, which
is consistent with the $\sim$ 2 reported in \citet{FUE00}.  We note a
spike in the ratio to values $>$ 3 near the starward edge of the
dissociation front which is possibly the result of the greater
ability of ortho-H$_{2}$ to self-shield \citep{DAB96}. This
measurement is not statistically significant, however, and is
consistent with an {\em R$_{OP}$} = 3. 

We observe a small but still significant
separation of ortho- and para-H$_{2}$ $\sim$ 2$\arcsec$, which is
shown in Figure
\ref{fig-orth_par}. Between 30-40$\arcsec$ from HD 200775 there is an
increase in {\em R$_{OP}$} which occurs at
the HI/H$_{2}$ transition.  While the detection of {\em R$_{OP}$} $>$ 3
in NGC 7023 is not statistically
significant, these profiles and
the {\em R$_{OP}$} spike suggest that we are observing
the effects of preferential self-shielding of ortho-H$_{2}$. If this
interpretation is correct, it would be
the first detection of the effects of preferential self-shielding of
ortho-H$_{2}$ in a PDR. 

The bright ribbon of NGC 7023 has been a
target of numerous H$_{2}$ studies from the far-UV \citep{FRA06,FRA09} to
the NIR \citep{LEM96,TAK00} and MIR \citep{FUE00}, however to our
knowledge this work is the first to derive properties of H$_{2}$
emission between the dissociation front and HD 200775, where the gas is subject to a
harsher UV radiation field and higher {\em G$_{o}$/n$_{H}$} (Figure
\ref{fig-7023a} (h)). Previous H$_{2}$ 1-0 S(1) and other narrow band imaging \citep{LEM96,TAK00} failed to detect any H$_{2}$ emission in this
region. \citet{FUE00} reported a weak 0-0 S(1) detection with {\em ISO}, however they did not detect any other rotational lines at
that location.  We find that the IRS SL band H$_{2}$ in this region is generally
hotter ($\sim$ 1000K), at nearly an order
of magnitude lower {\em N$_{H_{2}}$} ({\em N$_{H_{2}}$} $\sim$ 1 $\times$
10$^{19}$ cm$^{-2}$), and
close to an {\em R$_{OP}$} of 3. 

Previous MIR studies of NGC 7023 with {\em ISO} \citep{FUE00, RAPA05}
and {\em Spitzer} \citep{BRN07} have indicated a warm-H$_{2}$ at {\em T$_{gas}$}
$\sim$ 400 -- 700K and {\em N$_{H_{2}}$} $\sim$ 10$^{20}$
cm$^{-2}$. These values are not matched by those listed
in Table \ref{tbl-avgs}, however the data is the table is skewed by
giving equal weight to {\em T$_{gas}$} and {\em N$_{H_{2}}$}
calculated for the H$_{2}$ starwards of the bright emission ribbon. When the flux of each
rotational line over the entire observing area
is rebinned into a single value, which is dominated by flux from the ribbon, we calculate a {\em T$_{gas}$} $\sim$
650K, {\em N$_{H_{2}}$} $\sim$ 6 $\times$ 10$^{19}$ cm$^{-2}$, and
{\em R$_{OP}$} $\sim$ 1.83. These values are in line with previous work. 

\subsection{NGC 2023 North}
The PAH emission in NGC 2023N peaks $\sim$ 10 $\arcsec$ away from the
peak of the total detected SL band H$_{2}$ emission. The
separation is more pronounced between the PAHs and emission from the
para-H$_{2}$ states (including S(0)) than the
ortho-H$_{2}$ states (including S(1)). This can be seen by noting that
the {\em R$_{OP}$} peak in Figure \ref{fig-2023Na} (g) is
closer to the PAH emission peak in Figure \ref{fig-2023Na} (a) than
the H$_{2}$ peak in Figure \ref{fig-2023Na} (b). This indicates that
the direction of the advancing photodissociation front is moving from
the lower right of the maps towards the upper left, even though HD
37903 is located closer to the bottom left. 

\subsubsection{PAH Features}
The PAH ionization and dehydrogenation relations in Figures
\ref{fig-2023Na} (c) and (d) both reach a minimum near the PAH
emission peak and increase
with decreasing PAH emission. The low {\em I$_{7.7}$/I$_{11.3}$}
values indicate that the PAHs are primarily neutral throughout the
observed area. It is likely that the slight increase in PAH ionization
to the north of the dissociation front in NGC 2023N is due to a decreasing {\em
  n$_{H}$} and relatively flat {\em G$_{o}$}, therefore increasing the
ratio {\em G$_{o}$/n$_{H}$} as seen in Figure \ref{fig-2023Na} (h). 

\subsubsection{H$_{2}$ Features}
NGC 2023N is also a well studied PDR, with observations from the FUV \citep{BGH02,FRA09} to the NIR
\citep{HAS87,GATLEY87,TAK00}.    Unlike NGC 7023, we calculate a nearly flat
and featureless gas temperature and only
a factor of $\sim$ 2 drop in {\em N$_{H_{2}}$} outside of the
H$_{2}$ bright ribbon (Figures \ref{fig-2023Na} (f) and (e)). When all
of the SL band H$_{2}$ is summed, as in Figure \ref{fig-2023Na} (b),
the para-H$_{2}$ dominates the flux and {\em N$_{H_{2}}$}. 

The separation between ortho- and para-H$_{2}$ is larger in
NGC 2023N than in NGC 7023. Figure \ref{fig-2023orth_par} shows a $\sim$
10$\arcsec$ separation between the ortho- and para-H$_{2}$ peaks which
exists not only in the SL band rotational lines (S(2) -- S(5)), but
also in the LL rotational lines S(0) and S(1).  NGC 2023N is not oriented so that the
projected vector towards the central star is perpendicular to the dissociation
front, nevertheless Figures \ref{fig-2023Na} (f), (g),
and Figure \ref{fig-2023orth_par} show that {\em R$_{OP}$} is nearly identical on
either side of the {\em R$_{OP}$} peak and {\em T$_{gas}$} is flat
throughout. This implies that collisional conversion has
transitioned the {\em R$_{OP}$} from a formation value to $\sim$ 1.25 ({\em R$_{OP}$} calculated
using only the S(0) and S(1) lines for {\em T$_{gas}$} = 200 K is
0.67), and that the passing of the dissociation front is causing a temporary peak in
{\em R$_{OP}$} by preferential self-shielding of the ortho-H$_{2}$
in both the cool, diffuse gas and the warm, dense clumps
simultaneously. 

\subsection{NGC 2023 South}
NGC 2023S contains a bright emission ridge $\sim$ 78$\arcsec$
south of HD 37903 which has been thoroughly studied from the far-UV \citep{BGH02}
to the radio \citep{STI97}
and has been used as a template for modeling of dissociation fronts
\citep{DAB96}. The ridge area is considered to be denser ($\sim$
10$^{5}$ cm$^{-3}$) and have a slightly more thermal H$_{2}$
vibrational spectrum than its northern counterpart. The NGC 2023S IRS
coverage extends over a significantly broader area than the bright
emission ridge, however, extending up to $\sim$ 15$\arcsec$ from HD 37903 to as
far as 100$\arcsec$ away.

\begin{figure}
\plotone{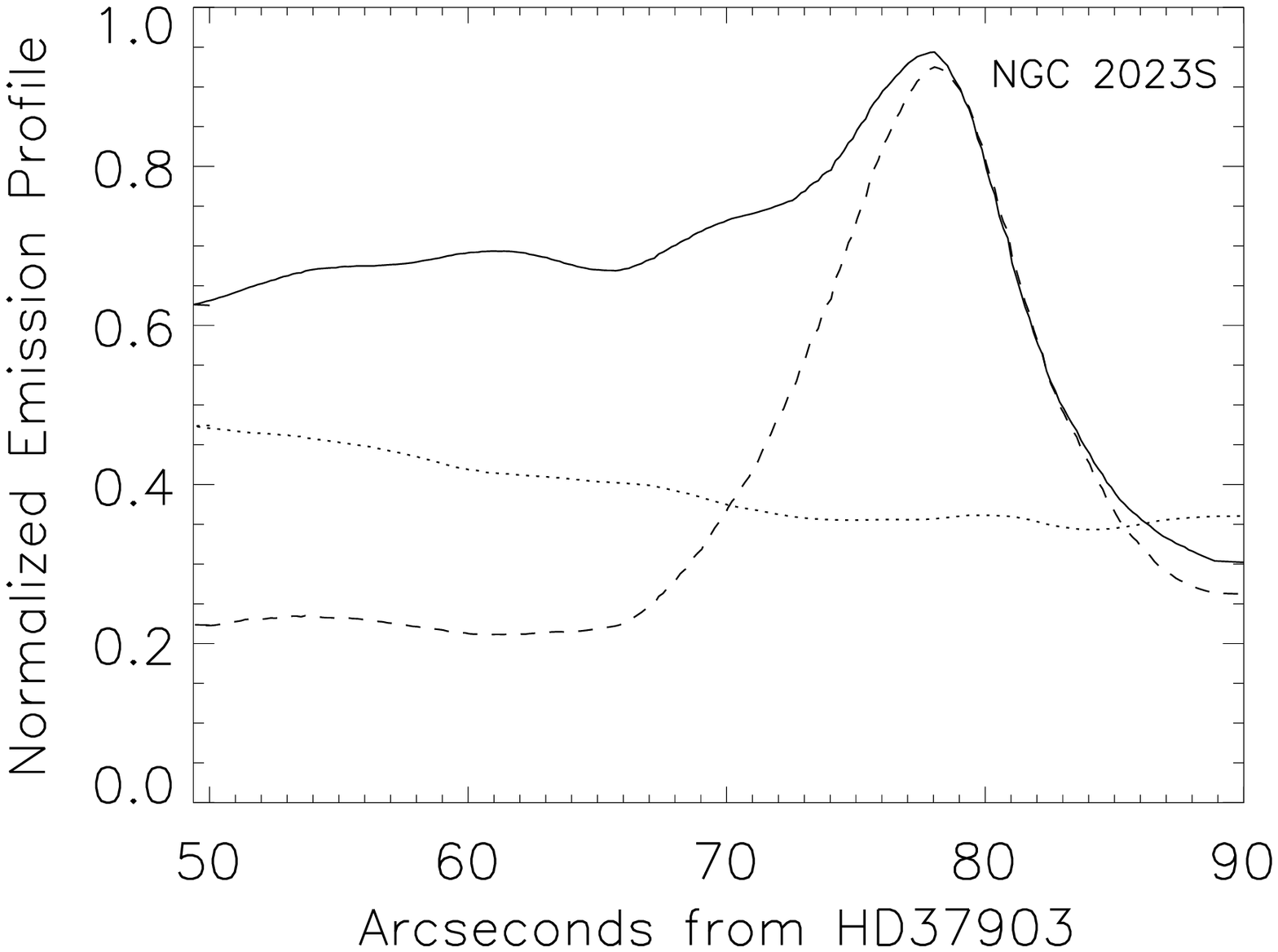}
\caption{Average normalized emission profile for PAHs (solid), H$_{2}$
  (dash) and (I$_{7.7}$/I$_{11.3}$)/12 (dot) for NGC 2023S. Multiple
  cross-sections through the PDR originating from the direction of HD 37903 are averaged to produce this profile.  
\label{fig-2023_2cut}}
\end{figure}

The H$_{2}$ and PAH emission in Figures \ref{fig-2023Sa} (a) and (b)
both are at a maximum in the bright emission ridge centered at
approximately [0,-20]. As with NGC 7023, the PAH emission remains nearer
to its peak flux level at smaller projected distances to the central star than the H$_{2}$. Figure \ref{fig-2023_2cut} shows that the
PAH emission peak precedes the H$_{2}$ emission peak in the bright
ridge, as it did in NGC 2023N and NGC 7023, but by only $\sim$
0.6$\arcsec$, which is not statistically significant. Higher resolution
measurements would be needed to confirm this separation. 

\subsubsection{PAH Features}
The PAH ionization band ratio {\em I$_{7.7}$/I$_{11.3}$} in the bright
emission ridge is at a similar average value ({\em
  I$_{7.7}$/I$_{11.3}$} $\sim$ 3, {\em f$_{+} \sim$} 0.25) to the ratio in
the ribbon of NGC 7023. To the north of the emission ridge the ratio
increases to a maximum value of just less than 7, indicating an
ionization fraction of {\em f$_{+} \sim$} 0.65. In the mapped region
closest to HD 37903, however, the PAH ionization level drops, possibly
due to the complex geometry of NGC 2023S resulting in foreground or
background contributions to the PAH emission. 

The PAH dehydrogenation band ratio map (Figure \ref{fig-2023Sa} (d))
may be contaminated by the presence of a bright star at the bottom
edge of the mapped area. The star and another fainter one to the north
were removed to avoid contamination, however at 12.7 $\mu$m some
contamination still exists. The concentric rings around the blacked
out area in Figure \ref{fig-2023Sa} (d) suggest that this may be
affecting the measurement. In the northern part of the map we observe
values similar to NGC 7023 and NGC 2023N with slightly higher ratios
nearer to the central star. 

\subsubsection{H$_{2}$ Features}
The bright emission ridge coincides with the highest {\em N$_{H_{2}}$}
in NGC 2023S ({\em N$_{H_{2}}$} $\sim$ 1.87 $\pm$ 0.30 $\times$
10$^{20}$ cm$^{-2}$, Figure \ref{fig-2023Sa} (e)). Within this region the
temperature averages $\sim$ 625 $\pm$ 13K
with an {\em R$_{OP}$} $\sim$ 1.87 $\pm$ 0.1. The spread on the values
cited represent the standard deviation within the region, not the
error on the measurements. We cite standard deviations here to show
the relative homogeneity of of the observations within the regions
discussed in this section. Farther to the south of the ridge,
{\em N$_{H_{2}}$} drops off slightly to
$\sim$ 8.9 $\pm$ 3.5 $\times$ 10$^{19}$ cm$^{-2}$, while {\em T$_{gas}$} (592 $\pm$ 32K) and
{\em R$_{OP}$} (1.7 $\pm$ 0.2)
remain at levels similar to those within the emission ridge. In the top left portion of the maps, nearest to HD 37903 and
coincident with the area where we noted a drop in {\em
  I$_{7.7}$/I$_{11.3}$} in $\S$4.3.1, {\em N$_{H_{2}}$}, {\em
  T$_{gas}$} and {\em R$_{OP}$} are all consistent with the values in
the southern portion of the map ({\em T$_{gas}$}  $\sim$ 575 $\pm$ 42K, {\em
  N$_{H_{2}}$} $\sim$ 6.9 $\pm$ 1.6 $\times$ 10$^{19}$ cm$^{-2}$, {\em R$_{OP}$}
$\sim$ 1.63 $\pm$ 0.41). The similarity in these properties suggests that the gas dominating the
emission in the upper left may be a foreground or background portion
of NGC 2023S enveloping the central cavity.  

North-west of the dissociation front, {\em N$_{H_{2}}$} drops by over an
order of magnitude relative to the ridge feature ({\em N$_{H_{2}}$}
  $\sim$ 1.9 $\pm$ 1.4 $\times$ 10$^{19}$ cm$^{-2}$). The corresponding
increase in {\em T$_{gas}$} to $\sim$ 847 $\pm$ 325K suggests that the H$_{2}$ in this region
of NGC 2023S is similar to the H$_{2}$ starwards of the dissociation
front in NGC 7023. The increased PAH ionization fraction and higher
{\em G$_{o}$/n$_{H}$} in Figure \ref{fig-2023Sa} (h) also follow the
PAH behavior starwards of the dissociation front in NGC 7023. {\em
  R$_{OP}$}, however, rather than increasing to the LTE {\em R$_{OP}$} $\sim$ 3,
falls to extremely low values $\sim$ 0.6 $\pm$ 0.22. This unusual observation will
be discussed further in $\S$ 5.3.1.

\subsection{IC 63}
\begin{figure}
\epsscale{0.79}
\plotone{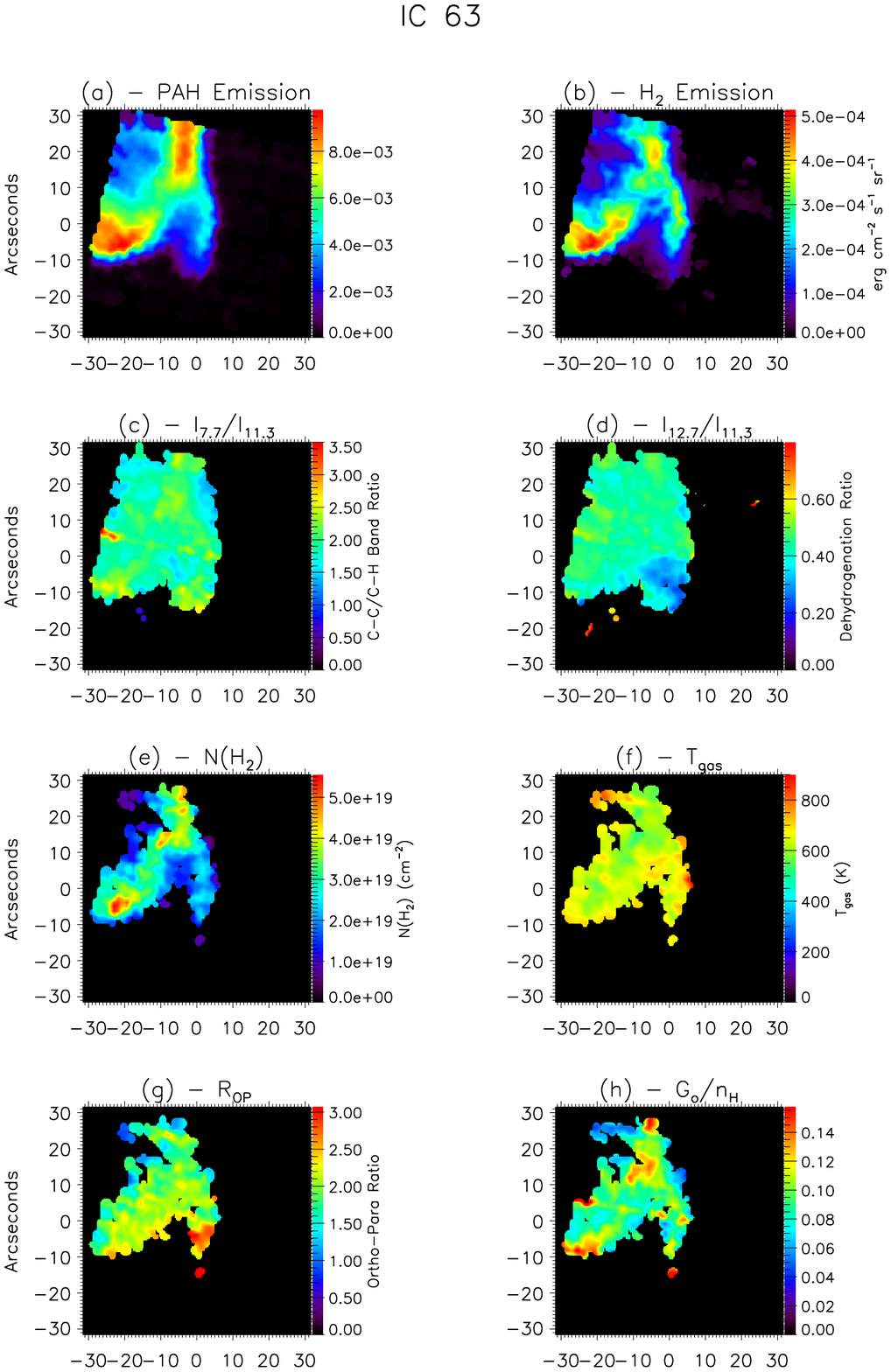}
\caption{PAH and H$_{2}$ emission and diagnostics for IC 63. (a)
  Total detected PAH emission. (b) Total detected H$_{2}$
  emission. (c) {\em I$_{7.7}$/I$_{11.3}$}. (d) {\em
    I$_{12.7}$/I$_{11.3}$}. (e) H$_{2}$ column denisty. (f) H$_{2}$
  derived gas temperature. (g) H$_{2}$ Ortho-to-para ratio. (h) {\em G$_{o}$/n$_{H}$}.
\label{fig-63a}}
\end{figure}

\begin{figure}
\plotone{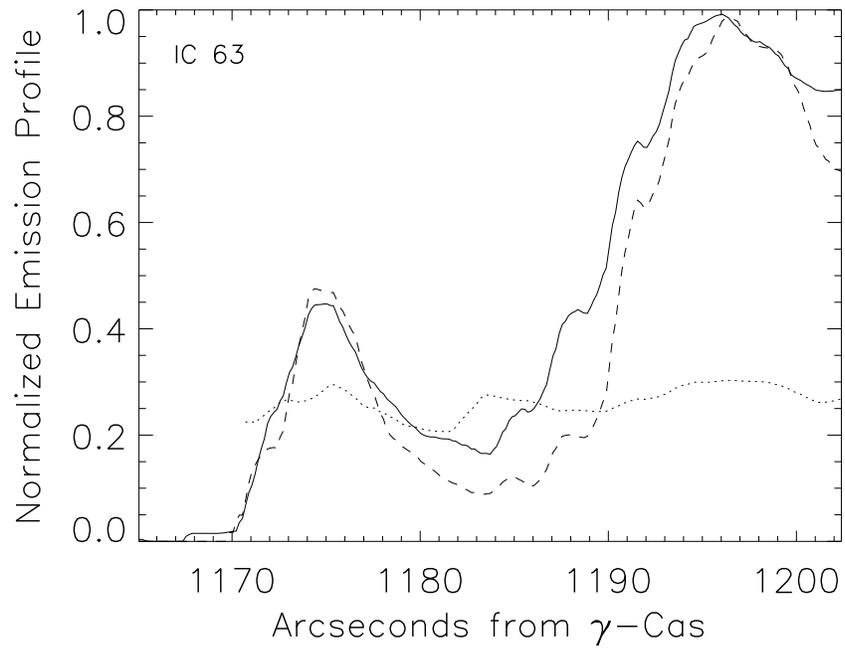}
\caption{Average normalized emission profile for PAHs (solid), H$_{2}$
  (dash) and (I$_{7.7}$/I$_{11.3}$)/12 (dot) for IC 63. Multiple
  cross-sections through the PDR originating from the direction of $\gamma$-Cas are averaged to produce this profile.  
\label{fig-63cut}}
\end{figure}

IC 63 features the weakest PAH and H$_{2}$ flux in our sample and the
lowest ratio of PAH flux to H$_{2}$ flux (Figure \ref{fig-63a} (a) and
(b)). We detect no statistically significant separation of the PAH and H$_{2}$
emission in IC 63 (Figure \ref{fig-63cut}). We also calculate the
lowest average {\em I$_{7.7}$/I$_{11.3}$} ratio in our sample with an
estimate ionization fraction of {\em f$_{+} \sim$} 0.082. This ratio
does not reach the average values observed in the other PDRs anywhere in the observed region (Figure
\ref{fig-63a} (c)).  It is worth noting that IC 63 is at a larger
distance from the exciting star $\gamma$-Cas ($\sim$ 1.3 pc) than NGC
2023 ($\sim$ 0.33 pc to NGC 2023N) and NGC
7023 ($\sim$ 0.1 pc), therefore {\em G$_{o}$} is lower due to
geometric effects despite the harder radiation spectrum from the B0 star. Figure \ref{fig-63a} (d) shows that the
dehydrogenation relation {\em I$_{12.7}$/I$_{11.3}$} is mostly
constant, however there is a decrease in the ratio near
the starward edge of the PDR. We will investigate this further in
$\S$5.1.2.

The temperature, column density, and {\em G$_{o}$/n$_{H}$} all
fluctuate only moderately (Figures \ref{fig-63a} (e), (f) and
(h)). The average measured {\em T$_{gas}$} of $\sim$ 625K is
consistent with the temperature cited in \citet{HAB04} and
\citet{KARR05} ({\em T$_{gas}$} = 630K).  IC 63 was the subject of an extensive multiwavelength
study in \citet{KARR05} in which {\em ISO} MIR data was use to compute
temperatures and column densities, as well as to compare PAH band
ratios. They reported a nearly flat PAH {\em I$_{11.3}$/I$_{6.2}$}
band ratio, a probe of PAH ionization, which is in good
agreement with the low variation of {\em I$_{7.7}$/I$_{11.3}$} shown
in Figure \ref{fig-63a} (c). They calculate an average {\em I$_{11.3}$/I$_{6.2}$} $\sim$ 2, however, which is twice the average
that we calculate for this PDR ({\em I$_{11.3}$/I$_{6.2}$} $\sim$
0.93), and would indicate even less PAH ionization if correct. The {\em R$_{OP}$} in Figure \ref{fig-63a} (g) peaks at a nearly LTE
value at the starwards edge of IC 63 before moderating down to $\sim$
1.5 at the back edge of our observing region. This implied trend in {\em R$_{OP}$} can only be confirmed with a
more extended coverage area, however it is consistent with an
advancing photodissociation front.

\section{Discussion}
In this section, we address connections that can be made between the
properties that we
have observed in these PDRs. We address the limited understanding of
the dehydrogenation relation and reference a proposed H$_{2}$
formation scenario which may explain the observed band variations. We
also consider whether our calculation of {\em G$_{o}$/n$_{H}$} is a
valid one given that the dominant H$_{2}$ emission in the IRS SL bandpass is from warm, dense gas clumps whereas the dominant
PAH emission may not be. We conclude by analyzing our ortho-to-para
ratio maps and how observed trends fit with current understanding of
advancing dissociation fronts. 

\subsection{PAH Dehydrogenation}
\subsubsection{NGC 7023 and NGC 2023}
\begin{figure}
\plotone{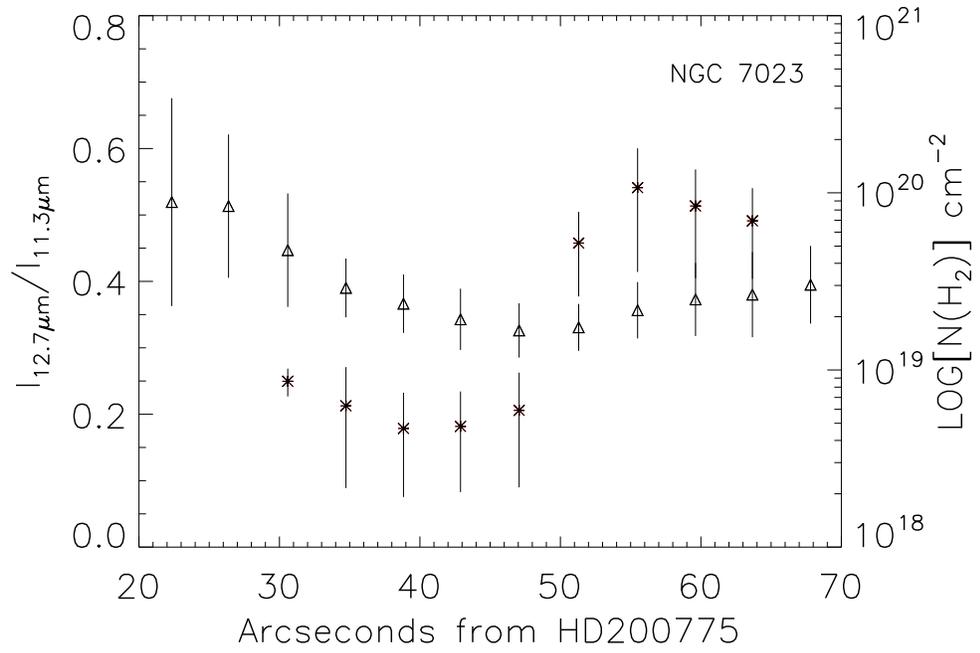}
\caption{Plot of the average profiles of {\em I$_{12.7}$/I$_{11.3}$}
  (triangle) and {\em N(H$_{2}$)} (asterix) versus
  distance to HD 200775 in NGC 7023.
\label{fig-dehy_nh}}
\end{figure}

In NGC 7023, NGC 2023N, and NGC 2023S,
  we observe an increase in the PAH dehydrogenation relation, {\em
    I$_{12.7}$/I$_{11.3}$}, beginning within the bright
  H$_{2}$ ribbons and increasing slightly with greater distance from
  the central stars. This increase indicates that the PAHs are being stripped of hydrogen atoms
despite corresponding drops in the ionization relation, {\em
 I$_{7.7}$/I$_{11.3}$}, and {\em G$_{o}$/n$_{H}$} (Figures
\ref{fig-7023a}, \ref{fig-2023Na}, \ref{fig-2023Sa} (c), (h)). In all
three PDRs, {\em I$_{12.7}$/I$_{11.3}$} reaches a minimum, indicating
the peak of PAH hydrogenation, just starwards of the H$_{2}$ peak,
where the H$_{2}$ is dissociating. The {\em
I$_{12.7}$/I$_{11.3}$} relation then increases again nearest the
central stars in NGC 7023 and NGC 2023S. This effect is illustrated
in Figure \ref{fig-dehy_nh}, which shows the averaged profiles of {\em
  I$_{12.7}$/I$_{11.3}$} and {\em N(H$_{2}$)} with respect to HD
200775 in NGC 7023. The corresponding drop in
total PAH emission near HD 200775 seen in Figure \ref{fig-7023cut} may be evidence of the evaporating of PAH molecules
due to a higher {\em G$_{o}$/n$_{H}$} and ionization fraction.  This would be consistent with
previous results showing that PAH ionization precedes dissociation in
the reflection/emission nebula IC 405 \citep{FRA07}. 

The PAH dehydrogenation in the molecular layer may be
due to a different process. H$_{2}$ formation on PAH surfaces has been
proposed as a dominant formation process in high temperature PDRs \citep{HAB04},
overtaking the traditional ISM formation process between physisorbed H
atoms on grain surfaces, which is effective only at low
temperatures. \citet{LePage09} propose a formation
mechanism where an ionized, fully hydrogenated PAH encounters an H
atom and becomes overly hydrogenated (the gas phase hydrogen atom
becomes bound to the surface of the PAH) and then recombines with a
free electron, which dislodges either the extra H atom or an H$_{2}$
molecule. In the H$_{2}$ ejection scenario, the neutral PAH product would be left with one fewer
hydrogen atom than it had previously, and thus is dehydrogenated. The
observed increase in {\em I$_{12.7}$/I$_{11.3}$} may indicate the presence of
dehydrogenated PAHs due to an ongoing H$_{2}$ formation process.  

Neither PAH destruction nor H$_{2}$ formation fully explains the
decrease in {\em I$_{12.7}$/I$_{11.3}$} just starwards
of the dissociation front, which is most easily seen in NGC 7023
(Figure \ref{fig-7023a} (d)). This region has a higher PAH ionization
fraction than in the molecular layer, which is an important catalyst
for most H$_{2}$ formation scenarios on PAH surfaces
\citep{Bauschlicher98,Pauzat01,LePage03}, therefore we would expect
the formation rate to increase as more PAH cations become
available. If the dehydrogenation ratio in the molecular layer is due
to H$_{2}$ formation, then the lower {\em
  I$_{12.7}$/I$_{11.3}$} starwards of the dissociation front would indicate a quenching of that
process. 

\subsubsection{PAH Destruction in IC 63}
We detect no significant variation of {\em I$_{12.7}$/I$_{11.3}$} in
the emission nebula IC 63 other than a minimum near the southern
starwards edge (Figure \ref{fig-63a} (d)). This region corresponds the
highest {\em R$_{OP}$} in IC 63, suggesting selective self-shielding of ortho-H$_{2}$
and therefore an HI/H$_{2}$ transition
layer. A dehydrogenation minimum near the leading edge of the
HI/H$_{2}$ transition is consistent with observations in NGC 7023 and
NGC 2023. IC 63 also exhibits FUV H$_{2}$ fluorescence indicative of very
little extinction between the PDR and $\gamma$-Cas
\citep{WITT89,FRA05} and the PAH emission region has been shown to be
enveloped in an ionized atomic layer $\sim$ 0.002 pc thick
\citep{KARR05,FRA05}. That the PAH emission does not continue into the HII
region traced by H$\alpha$ indicates that the PAH molecules are being
dissociated near the starwards edge of the PDR. Figure \ref{fig-63a}
(d), however, does not show any evidence of an increase in the dehydrogenation
relation, suggesting that either that the dissociation is
happening too quickly to be resolvable or that the dehydrogenation signature is too
weak to detect. PAH molecules can be dissociated directly by the absorption
of single high energy photons, however the required photon energy
increases with the number of carbon atoms in the core, with a PAH of
{\em N$_{C}$} = 50 requiring an EUV photon of $\sim$ 24 eV
\citep{Sieben10}. Recent studies have shown that $\gamma$-Cas is a
soft x-ray emitter \citep{LOPES10}, making single photon dissociation
a possible scenario. 

The average value of the {\em I$_{12.7}$/I$_{11.3}$} dehydrogenation
ratio in IC 63 is similar to the average in the H$_{2}$ bright ribbons
of NGC 2023 and NGC 7023 ($\sim$ 0.41), suggesting
that the same processes that influence this ratio are occurring in all four
PDRs. The ionization relation however (Figure \ref{fig-63a} (c)),
indicates that IC 63 is very close to the PAH neutral limit of {\em
  I$_{7.7}$/I$_{11.3}$} $\sim$ 1.3, limiting the number of PAH cations
available for H$_{2}$ formation. 

\subsection{Reconciliation of {\em G$_{o}$/n$_{H}$}}
To calculate {\em G$_{o}$/n$_{H}$} we have used the PAH 7.7 $\mu$m and 11.3
$\mu$m flux in conjunction with the temperature of the warm H$_{2}$
component, which represents $\sim$ 1$\%$ of the total H$_{2}$
column (Figure \ref{fig-tempcurv}). This method has been used before
\citep{GAL08,BRN09}, however it assumes that the PAH ionization
relation is independent of the local hydrogen density. In our
example, where we have used the warm H$_{2}$ temperature to calculate
{\em G$_{o}$/n$_{H}$}, we are assuming that we are calculating the
ratios within those dense gas clumps. 

Radio observations of PDRs often use other molecules, such as CO, as a
proxy for H$_{2}$ \citep{CKS88}. These studies predict H$_{2}$
temperatures for the low density, cool H$_{2}$ component ({\em
  n$_{H}$} $\sim$ 10$^{3}$ cm$^{-3}$, {\em T$_{gas}$} $\sim$ 100K),
while MIR studies of the warm, dense H$_{2}$ clumps have indicated densities
$\sim$ 10$^{4}$ -- 10$^{5}$ cm$^{-3}$ and temperatures $\sim$
600K. For these cases, the difference in temperature, $\sqrt{T_{warm}/T_{cool}}
\sim$ 2.5, does not account for the order of magnitude or more
difference in {\em n$_{H}$} using the equations in $\S$3.4. 

In order for the actual value calculated to be the correct {\em G$_{o}$/n$_{H}$} within the
warm, dense H$_{2}$ clumps that we are studying, the PAH flux must be
dominated by emission from within those same high density clumps,
thereby probing the same environment as the warm H$_{2}$
temperature that we are using. If this is not the case, then either {\em G$_{o}$} is
larger within the high density clumps than in the low density, cool
component, or we are improperly using the warm {\em T$_{gas}$} and
obtaining an incorrect measure of {\em G$_{o}$/n$_{H}$}. We therefore
caution that our maps of {\em G$_{o}$/n$_{H}$} be used primarily as a
measure of the variation of the ratio over the PDR, and not as a means
of obtaining an absolute value for a specific portion of the PDR. 

\subsection{H$_{2}$ Ortho-Para Ratio}
The formation ratio of ortho- to para-H$_{2}$ ({\em R$_{OP}$}) in a warm gas (T$_{gas}$
$\geq$ 250 K) is expected to be 3. H$_{2}$ that forms
at a cooler temperature will have a lower {\em R$_{OP}$} as more
H$_{2}$ is formed in the {\em J} = 0 para-rotational state. Subsequent
radiative heating of the gas allows {\em J} to change by only $\pm$ 2, thus the formation {\em R$_{OP}$} is
preserved in higher {\em J} rotational states. Radiative
processes cannot convert ortho-H$_{2}$ to para-H$_{2}$, however the total nuclear spin can be altered by collisions with H and H$^{+}$
atoms, which moves {\em R$_{OP}$} towards the local thermal
equilibrium value defined by the new temperature \citep{SNU99}. 
 
The observed {\em R$_{OP}$} for a temperature, {\em T$_{gas}$}, and density of atomic
hydrogen, {\em n$_{HI}$}, after a time $\tau$ can be predicted by
Equation [2] from \citet{NEU06}.  Assuming a {\em T$_{gas}$} $\sim$
650K, a density of {\em n$_{H}$} $\sim$ 10$^{5}$, {\em n$_{HI}$} =
10$^{-3}$n$_{H}$, and an initial ortho-to-para ratio of {\em R$_{OP_o}$} = 0.75
(corresponding to a formation temperature $\sim$ 50 - 75K),
the observed {\em R$_{OP}$} after 5000 years at {\em T$_{gas}$}
should be $\sim$ 2.15. This is within the 1$\sigma$ errors of the average
{\em R$_{OP}$} we measure for the ribbon of NGC 7023 (1.86 $\pm$ 0.32), which has been exposed
to HD 200775 for $\sim$ 5000 years \citep{ROG95}. It is also
consistent with IC 63 ({\em R$_{OP}$} = 1.99 $\pm$ 0.37) which has
similar {\em T$_{gas}$} and {\em n$_{H}$}. 

A PDR with a moving dissociation front should exhibit a gradient in both the temperature and
{\em R$_{OP}$} as the front advances into the cool, low {\em R$_{OP}$}
gas, heats it, and begins the para $\rightarrow$ ortho conversion. This process
is most clearly observed in NGC 7023 in Figures \ref{fig-7023a} (f)
and (g), where the cool, outer regions of the PDR have a lower {\em R$_{OP}$} than nearer
to the central star, where {\em R$_{OP}$} is closer to 3. It is also apparent in Figure \ref{fig-orth_par} where beyond the H$_{2}$
emission peak {\em R$_{OP}$} falls to $\sim$ 0.6, reflecting a low formation
temperature, while {\em T$_{gas}$} levels out, indicating recent heating.  We also observe a spike in {\em
  R$_{OP}$} at the starwards edge of the dissociation front in NGC
2023N, however the measured values do not exceed 3. There is also the possibility that
the {\em R$_{OP}$} $\sim$ 3 closest to HD 200775 in NGC 7023 is the
result of rapid dissociation and formation processes and the
relatively small amount of H$_{2}$ detected in that region is newly
formed H$_{2}$, not relic H$_{2}$ which has transitioned to LTE. 

All four PDRs studied in this paper exhibit non-equilibrium ortho-to-para
ratios which indicate that the H$_{2}$ formed at much lower
temperatures prior to illumination by the central stars.  The ortho-to-para ratios in IC 63 and NGC 7023 generally follow
expectations for a recently heated PDR, however more coverage area in
IC 63 is needed to confirm the implied trends in Figure \ref{fig-63a} (g). 

\subsubsection{Extremely Low {\em R$_{OP}$} in NGC 2023 South}
In NGC 2023S we calculate very low non-equilibrium ortho-to-para ratios within a pocket of warmer, low
N(H$_{2}$) gas at a smaller projected distance to the central star than the bright emission
ridge (Figure \ref{fig-2023Sa} (g)).  Ratios in this region, where the temperature occasionally
exceeds 1000K, fall to $<$ 1 with an average of 0.6 $\pm$ 0.22. At such high temperatures the
para $\rightarrow$ ortho conversion process should bring the gas to
almost LTE within 1000 years and to {\em R$_{OP}$} $>$ 1 within a few hundred years for
n$_{H} \sim$ 5 $\times$ 10$^{4}$ cm$^{-3}$ \citep{NEU06}. We know of only two scenarios which could explain such extreme values. Either the gas
has only been recently and rapidly heated, or the density of the gas is
significantly lower than in the rest of the PDR, slowing the
transition process. This region coincides with the highest {\em G$_{o}$/n$_{H}$} in NGC
2023S (Figure \ref{fig-2023Sa} (h)), implying either a strong UV
radiation field or a low density. Such low ortho-to-para ratios
for warm (T$_{gas}$ $\geq$ 500K) have been reported in \citet{NEU06}
and \citet{Maret09}, however only for gas
recently heated by a passing shock wave. This region merits further
study to better understand and confirm these results. 

\section{Conclusions}
In the photodissociation regions near hot stars, the MIR
spectrum is dominated by emission from PAHs and the pure rotational lines of H$_{2}$. We
have shown that while these two species are independent of each other,
they provide complementary diagnostic tools to probe the environment of the PDR. Utilizing the
spatial and spectral capabilities of the {\em Spitzer Space Telescope}, we
have created spectral maps of the ortho-to-para
ratio, temperature, column density of H$_{2}$, {\em
  G$_{o}$/n$_{H}$}, PAH ionization, and a measure of the PAH surface structure. These diagnostics, combined with spectral mapping
techniques, enable the study of PDR environments on angular scales of
a few arcseconds. Such maps will prove valuable to future spectral mapping studies with $Herschel$, which does not have direct access
to the bulk of the PDR gas mass (H$_{2}$) in its far-infrared and sub-mm
bandpasses. 

\subsection{PAH Features}
Using the PAH emission in the MIR, we were able to establish an estimate
of {\em $\sqrt{T_{gas}}$G$_{o}$/n$_{H}$} across the entire coverage
  region using the {\em I$_{7.7\mu m}$/I$_{11.3\mu m}$} C-C/C-H band ratio. There is evidence for PAH dehydrogenation, and possibly destruction,
in NGC 7023 at radii $<$ 30$\arcsec$ ($\sim$ 0.058 pc) from HD 200775. The onset of PAH dehydrogenation in NGC 7023 coincides
with a drop in PAH intensity in every band and an ionization fraction of
{\em f$_{+}$} $\sim$ 0.74.  At projected distances $>$ 30$\arcsec$ the
dehydrogenation ratio reaches a minimum before increasing again in
the H$_{2}$/PAH bright ribbon feature. The dehydrogenation
signature in the molecular layer may indicate that H$_{2}$ is forming
on the PAH surfaces, and that the process is removing a bound hydrogen
atom from the PAH. We find that in
all three PDRs, {\em I$_{12.7}$/I$_{11.3}$} peaks where {\em T$_{gas}$}
$\sim$ 600-700K, possibly indicating the peak efficiency temperature
for H$_{2}$ formation on PAH grain surfaces. 

In three of the four PDRs studied, the PAH emission peaks a few
arcseconds closer to the central star than the H$_{2}$ emission.  The
projected separation distance varied from $\sim$ 0.6$\arcsec$ and
1.5$\arcsec$ in NGC 2023S and NGC 7023 respectively, to $\sim$
10$\arcsec$ in NGC 2023N. In IC 63, no separation was observed between H$_{2}$ and PAH
features, however the weak {\em G$_{o}$/n$_{H}$} of IC 63 or the
viewing geometry may
make the separation unresolvable. In NGC 2023N we observed no
significant separation between the PAHs and the ortho-H$_{2}$, however
the para-H$_{2}$ dominates the total H$_{2}$ emission and column
density in this PDR. This may also be the case in NGC 7023 where an
ortho- and para-H$_{2}$ separation was observed of similar magnitude
($\sim$ 2$\arcsec$) to the PAH-H$_{2}$ total separation. We lack the
resolution to definitively determine if the ortho-H$_{2}$ is cospatial
with the peak PAH emission in this PDR. No
separation between ortho- and para-H$_{2}$ was detected in NGC 2023S,
although there is a suggestion of a PAH-H$_{2}$ separation of $\sim$
0.6$\arcsec$, although it is not statistically significant. Higher
spatial resolution measurements are required to confirm a separation.  

\subsection{H$_{2}$ Features}
Despite the strong PAH features in these PDRs, we were able to
construct spectral line maps of multiple H$_{2}$ emission lines. We fit gas temperatures of the warm H$_{2}$ component
which closely match global H$_{2}$ derived temperatures from previous
studies, but with the added benefit of spatial resolution.
Due to the strength of the PAH features, we were unable to accurately
fit the H$_{2}$ S(4) and S(6) rotational lines in NGC 2023 and NGC
7023, however we estimate that the error introduced to the S(4) flux
is negligible in IC 63. For the purposes of temperature and column density calculations we applied a correction to the
S(4) flux to increase our spatial coverage area.  

In the bright emission ridges of NGC 7023 and NGC 2023N the
ortho-rotational emission peaks before the para-rotational
H$_{2}$; an effect most pronounced in NGC 2023N and not detected in
IC63 and NGC 2023S. In NGC 2023N,
where we had complete complementary IRS LL data, the peak of the
ortho-to-para ratio calculated for the cool, low density H$_{2}$, defined
by the S(0) and S(1) emission lines, is
cospatial with the peak of the ortho-to-para ratio for the warm, dense
clumps; suggesting that H$_{2}$ self-shielding occurs
at the same location regardless of the density or temperature
phase of the H$_{2}$ gas. 

\begin{figure}
\epsscale{0.79}
\plotone{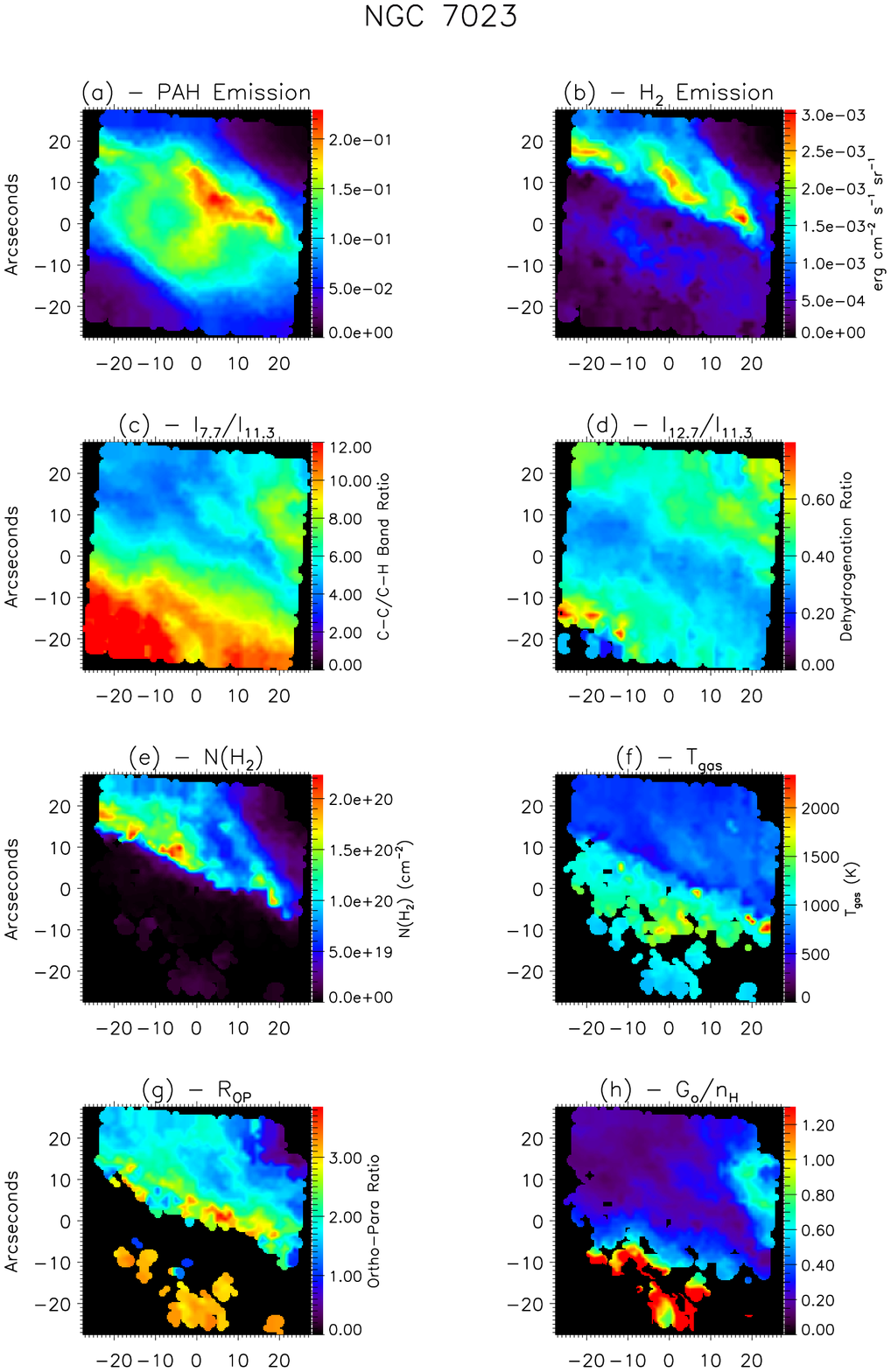}
\caption{PAH and H$_{2}$ emission and diagnostics for NGC 7023. (a)
  Total detected PAH emission. (b) Total detected H$_{2}$
  emission. (c) {\em I$_{7.7}$/I$_{11.3}$}. (d) {\em
    I$_{12.7}$/I$_{11.3}$}. (e) H$_{2}$ column denisty. (f) H$_{2}$
  derived gas temperature. (g) H$_{2}$ Ortho-to-para ratio. (h) {\em G$_{o}$/n$_{H}$}.
\label{fig-7023a}}
\end{figure}

\begin{figure}
\plotone{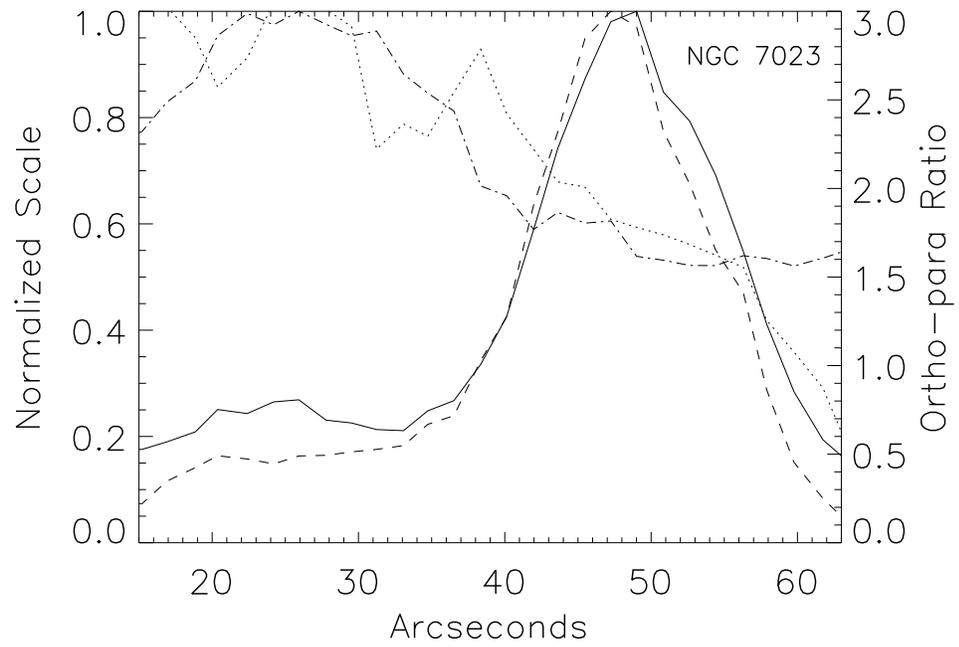}
\caption{Plot of the ortho-H$_{2}$ S(3) (dash) and para-H$_{2}$ S(2) (solid)
  emission in NGC 7023 as a function of projected distance from the
  central star. Also plotted is the average normalized temperature
  profile (dash-dot) and the ortho-para ratio (dot).    
\label{fig-orth_par}}
\end{figure}

\begin{figure}
\plotone{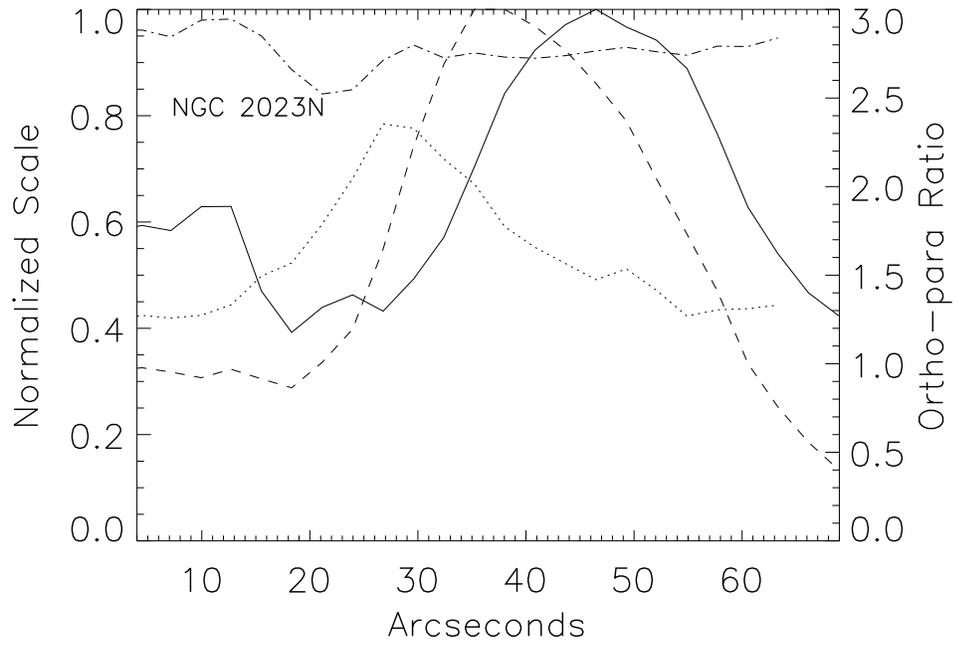}
\caption{Plot of the total ortho-H$_{2}$ (dash) and para-H$_{2}$ (solid)
  emission in NGC 2023N as a function of projected distance from the
  lower right corner of the maps in Figure \ref{fig-2023Na}. Also
  plotted is the average normalized temperature
  profile (dash-dot) and the ortho-para ratio (dot).    
\label{fig-2023orth_par}}
\end{figure}

We presented evidence of the preferential self-shielding of
ortho-H$_{2}$ in both NGC 7023 and NGC 2023N. We also detected an
implied trend in the ortho-to-para ratio in IC 63, with near LTE
values at the leading edge of the PDR falling to $\sim$ 1.5 at the
edge of our coverage area. In NGC 7023 we measured {\em R$_{OP}$} $>$ 3 (Figure \ref{fig-7023a} (g)), however the measurements
are not statistically significant and are consistent with the LTE
value of {\em R$_{OP}$} = 3. Ortho- and para-H$_{2}$ emission
profiles in NGC 2023N and NGC 7023 (Figures \ref{fig-2023orth_par} and \ref{fig-orth_par})
show that in both PDRs the ortho-H$_{2}$ emission peaks starwards of
the para-H$_{2}$ emission profile, which supports a preferential
self-shielding scenario. In NGC 2023N, we observe no lasting effect of preferential
self-shielding of ortho-H$_{2}$ on {\em R$_{OP}$}, as {\em R$_{OP}$} was nearly identical on either side of the dissociation
front. The ortho-to-para ratio of the cool H$_{2}$ component in NGC 2023N
is consistent with an initial formation temperature of 50 - 75K.

\acknowledgments
The authors would like to thank Dr. David Neufeld for helpful discussions
and advice. We would also like to thank the anonymous referee for
their helpful comments. This work is based in part on archival data obtained with
the {\em Spitzer Space Telescope}, which as operated by the Jet
Propulsion Laboratory, California Institute of Technology under a
contract with NASA. Support for this work was provided by an award
issued by JPL/Caltech for proposal ID: 30696.

\clearpage
\bibliography{flem0623}
\clearpage

\begin{deluxetable}{cccccccc}
\tabletypesize{\scriptsize}
\tablecaption{Literature Properties of Sample PDRs\label{tbl-prop}}
\tablewidth{0pt}
\tablehead{
\colhead{} & \colhead{NGC 7023} & \colhead{NGC 2023N (pos.1)} &
\colhead{NGC 2023S (pos.2)} & \colhead{IC 63} \\
}
\startdata
Central Star & HD 200775 & HD 37903 & HD 37903 & $\gamma$-Cas \\
Spectral Type & B2.5 Ve\tablenotemark{a} & B1.5 V\tablenotemark{b} &
B1.5 V\tablenotemark{b} & B0.5 IVe\tablenotemark{c}\\
T$_{eff}$ (K)& 17000\tablenotemark{a} & 22000\tablenotemark{b} &
22000\tablenotemark{b} & 25000\tablenotemark{c} \\
T$_{gas}$ (K) & 400-800\tablenotemark{d} & 550 & 550 & 620\tablenotemark{c} \\
n$_{H}$ (cm$^{-3}$) & 10$^{3}$-10$^{6}$\tablenotemark{d} &
10$^{4}$-10$^{6}$\tablenotemark{e} &
10$^{3}$-10$^{6}$\tablenotemark{e} &  10$^{4}$-10$^{5}$\tablenotemark{c} \\
G$_{o}$ & 2400\tablenotemark{d} & 2200\tablenotemark{e}
& 3-4000\tablenotemark{e} & 650\tablenotemark{c}\\
\enddata

\tablenotetext{a}{\citet{ROG95}}
\tablenotetext{b}{\citet{DeB83}}
\tablenotetext{c}{\citet{JANSEN94,KARR05}}
\tablenotetext{d}{\citet{FUE00, LEM96, CKS88}}
\tablenotetext{e}{\citet{DAB96, STI97}}
\end{deluxetable}

\clearpage

\begin{deluxetable}{ccccccccc}
\tabletypesize{\scriptsize}
\tablecaption{Summary of {\it Spitzer} Observations\label{tbl-obs}}
\tablewidth{0pt}
\tablehead{
\colhead{Object} & \colhead{Obs.} & \colhead{Program} &
\colhead{Target} & \colhead{Target} & \colhead{Exposure} &
\colhead{Number of} &
\colhead{Background} & \colhead{Background} \\
\colhead{} & \colhead{Date} & \colhead{ID} &
\colhead{RA} & \colhead{DEC} & \colhead{Time (s)} & \colhead{exposures} &
\colhead{Type} & \colhead{ID} 
}
\startdata
NGC 7023 & Aug. 8, 2004 & 28 & 21:01:31 & +68:10:43 & 14.68 & 30 & Nearby
Obs. & 22 \\
NGC 2023N & Mar. 12, 2005 & 3512 & 05:41:35 & -02:12:41 & 14.68 & 21 &
Nearby Obs. & 172
\\
NGC 2023S & Oct. 8, 2007 & 30295 & 05:41:37 & -02:16:45 & 29.36 & 108
& SKY & -- \\
IC 63 & Jan. 7, 2005 & 3512 & 00:59:00 & +60:53:10 & 14.68 & 31 & Outrigger
& -- \\
\enddata

\end{deluxetable}

\clearpage


\begin{deluxetable}{ccccccccc}
\tabletypesize{\scriptsize}
\tablecaption{Average Properties for Regions with Calculated T$_{gas}$\label{tbl-avgs}}
\tablewidth{0pt}
\tablehead{
\colhead{Object} & \colhead{Distance} & \colhead{T$_{avg}$} & \colhead{N(H$_{2}$)$_{avg}$} &
\colhead{Avg.} & \colhead{Avg.} & \colhead{Ortho-para}\\
\colhead{  } &  \colhead{from
  Star (\arcsec)\tablenotemark{a}} & \colhead{(K)} &
\colhead{(10$^{20}$ cm$^{-2}$)} & \colhead{I$_{7.7}$/I$_{11.3}$}&
\colhead{G$_{o}$/n$_{H}$} & \colhead{Ratio}
}
\startdata
NGC 7023 & 41.9 $\pm$ 11.7  & 873 $\pm$ 273 &
0.54 $\pm$ 0.53 & 6.06 $\pm$ 1.74 & 0.39 $\pm$ 0.38 & 1.96 $\pm$ 0.67 \\
NGC 2023N & 174.8 $\pm$ 14.4  & 522 $\pm$ 35 &
0.94 $\pm$ 0.33 & 5.53 $\pm$ 0.84 & 0.32 $\pm$ 0.12 & 1.64 $\pm$ 0.54 \\
NGC 2023S & 67.0 $\pm$ 24.0  & 689 $\pm$ 149 &
0.58 $\pm$ 0.40 & 5.38 $\pm$ 1.00 & 0.27 $\pm$ 0.10 & 1.26 $\pm$ 0.57 \\
IC 63 & 1192 $\pm$ 8.5  & 637 $\pm$ 51 &
0.26 $\pm$ 0.10 & 3.23 $\pm$ 0.38 & 0.10 $\pm$ 0.02 & 1.99 $\pm$ 0.37 \\
\enddata

\tablecomments{$\pm$ values represent standard deviations of the
  values calculated, not the error}
\tablenotetext{a}{Projected distance}
\end{deluxetable}

\clearpage

\clearpage

\clearpage


\clearpage

\end{document}